\documentclass[longauth,traditabstract]{aa}  
\usepackage{amsmath}
\usepackage{graphicx}
\usepackage{txfonts}
\usepackage{amssymb}
\usepackage[]{natbib}
\usepackage{mathrsfs}
\usepackage[colorlinks=true, linkcolor=blue, citecolor=blue, urlcolor=blue]{hyperref}
\usepackage{caption}
\usepackage{subcaption}
\usepackage{graphics}

\begin{document}

   \title{Kepler Object of Interest Network}

   \subtitle{I. First results combining ground and space-based observations of Kepler systems with transit timing variations}

   \author{C. von Essen$^{1,2}$, A. Ofir$^{2,3}$, S. Dreizler$^2$,
     E. Agol$^{4,5,6,7}$, J. Freudenthal$^2$, J. Hern\'andez$^8$,
     S. Wedemeyer$^{9,10}$, V. Parkash$^{11}$, H. J. Deeg$^{12,13}$,
     S. Hoyer$^{12,13,14}$, B. M. Morris$^4$, A. C. Becker$^4$,
     L. Sun$^{15}$, S. H. Gu$^{15}$, E. Herrero$^{16}$,
     L. Tal-Or$^{2,18}$, K. Poppenhaeger$^{19}$, M. Mallonn$^{20}$,
     S. Albrecht$^{1}$, S. Khalafinejad$^{21}$, P. Boumis$^{22}$,
     C. Delgado-Correal$^{23}$, D. C. Fabrycky$^{24}$, R. Janulis$^{25}$,
     S. Lalitha$^{26}$, A. Liakos$^{22}$, \v{S}. Mikolaitis$^{25}$,
     M. L. Moyano D'Angelo$^{27}$, E. Sokov$^{28,29}$,
     E. Pak\v{s}tien\.{e}$^{25}$, A. Popov$^{30}$,
     V. Krushinsky$^{30}$, I. Ribas$^{16}$, M. M. Rodr\'iguez S.$^8$,
     S. Rusov$^{28}$, I. Sokova$^{28}$,
     G. Tautvai\v{s}ien\.{e}$^{25}$, X. Wang$^{15}$}
   \authorrunning{C. von Essen
     et al. (2017)} \titlerunning{KOINet: study of exoplanet
     systems via TTVs} \offprints{cessen@phys.au.dk}

   \institute{$^1$Stellar Astrophysics Centre, Ny Munkegade 120, 8000, Aarhus, Denmark\\ 
   $^2$Institut f\"{u}r Astrophysik, Georg-August-Universit\"{a}t G\"{o}ttingen, Friedrich-Hund-Platz\,1, 37077 G\"{o}ttingen, Germany\\
              \email{cessen@phys.au.dk}\\
   $^3$Department of Earth and Planetary Sciences, Weizmann Institute of Science, Rehovot, 76100, Israel\\
   $^4$Astronomy Department, University of Washington, Seattle, WA 98195\\
   $^5$Institut d'Astrophysique de Paris, 98 bis Boulevard Arago, Paris 75014, France\\
   $^6$Guggenheim Fellow\\
   $^7$Virtual Planetary Laboratory\\
   $^8$Instituto de Astronom\'ia, UNAM, Campus Ensenada, Carretera Tijuana-Ensenada km 103, 22860 Ensenada, B.C. M\'exico\\
   $^9$Rosseland Centre for Solar Physics, University of Oslo, P.O. Box 1029 Blindern, N-0315 Oslo, Norway\\
   $^{10}$Institute of Theoretical Astrophysics, University of Oslo, P.O. Box 1029 Blindern, N-0315 Oslo, Norway\\
   $^{11}$School of Physics and Astronomy, Monash Centre for Astrophysics (MoCA), Monash University, Clayton, Victoria 3800, Australia\\
   $^{12}$Instituto de Astrof\'\i sica de Canarias, C. V\'\i a L\'actea S/N, E-38205 La Laguna, Tenerife, Spain\\
   $^{13}$Universidad de La Laguna, Dept. de Astrof\'\i sica, E-38206 La Laguna, Tenerife, Spain\\
   $^{14}$Aix Marseille Univ, CNRS, LAM, Laboratoire d'Astrophysique de Marseille, Marseille, France\\
   $^{15}$Yunnan Observatories, Chinese Academy of Sciences, P.O.Box 110, Kunming 650216, Yunnan Province, China\\
   $^{16}$Institut de Ciències de l'Espai (IEEC-CSIC), C/Can Magrans, s/n, Campus UAB, 08193 Bellaterra, Spain\\
   $^{17}$Observatori del Montsec (OAdM), Institut d’Estudis Espacials de Catalunya (IEEC), Gran Capit`a, 2-4, Edif. Nexus, 08034 Barcelona, Spain\\
   $^{18}$School of Geosciences, Raymond and Beverly Sackler Faculty of Exact Sciences, Tel Aviv University, Tel Aviv, 6997801, Israel\\
   $^{19}$Astrophysics Research Centre, Queen's University Belfast, Belfast BT7 1NN, UK\\
   $^{20}$Leibniz-Institut f\"{u}r Astrophysik Potsdam, An der Sternwarte 16, D-14482 Potsdam, Germany\\
   $^{21}$Hamburg Observatory, Hamburg University, Gojenbergsweg 112, 21029 Hamburg, Germany\\
   $^{22}$Institute for Astronomy, Astrophysics, Space Applications and Remote Sensing, National Observatory of Athens, Metaxa \& Vas. Pavlou St., Penteli, Athens, Greece\\
   $^{23}$Dipartimento di Fisica e Scienze della Terra, Università degli Studi di Ferrara, via Saragat 1, 44122 Ferrara, Italy\\
   $^{24}$Department of Astronomy \& Astrophysics, University of Chicago, Chicago, IL 60637\\
   $^{25}$Institute of Theoretical Physics and Astronomy, Vilnius University, Sauletekio 3, 10257 Vilnius, Lithuania\\
   $^{26}$Indian Institute of Astrophysics, Koramangala II block, Bangalore 560034, India\\
   $^{27}$Instituto de Astronom\'ia, Universidad Cat\'olica del Norte, Av. Angamos 0610, Antofagasta, Chile\\
   $^{28}$Central Astronomical Observatory at Pulkovo of Russian Academy of Sciences, Pulkovskoje shosse d. 65, St. Petersburg, Russia, 196140\\
   $^{29}$Special Astrophysical Observatory, Russian Academy of Sciences, Nizhnij Arkhyz, Russia, 369167\\
   $^{30}$Kourovka Astronomical Observatory of Ural Federal University, Mira Str. 19, 620002\\
   } 

   \date{Received 18/12/2017; accepted}

\abstract{During its four years of photometric observations, the
  Kepler space telescope detected thousands of exoplanets and
  exoplanet candidates. One of Kepler's greatest heritages has been
  the confirmation and characterization of hundreds of multi-planet
  systems via Transit Timing Variations (TTVs). However, there are
  many interesting candidate systems displaying TTVs on such long time
  scales that the existing Kepler observations are of insufficient
  length to confirm and characterize them by means of this
  technique. To continue with Kepler's unique work we have organized
  the ``Kepler Object of Interest Network'' (KOINet), a multi-site
  network formed by several telescopes spread among America, Europe
  and Asia. The goals of KOINet are to complete the TTV curves of
  systems where Kepler did not cover the interaction timescales well,
  to dynamically prove that some candidates are true planets (or not),
  to dynamically measure the masses and bulk densities of some
  planets, to find evidence for non-transiting planets in some of the
  systems, to extend Kepler's baseline adding new data with the main
  purpose of improving current models of TTVs, and to build a platform
  that can observe almost anywhere on the Northern hemisphere, at
  almost any time. KOINet has been operational since March, 2014. Here
  we show some promising first results obtained from analyzing seven
  primary transits of \mbox{KOI-0410.01}, \mbox{KOI-0525.01},
  \mbox{KOI-0760.01}, and \mbox{KOI-0902.01} in addition to Kepler
  data, acquired during the first and second observing seasons of
  KOINet. While carefully choosing the targets we set demanding
  constraints about timing precision (at least 1 minute) and
  photometric precision (as good as 1 part per thousand) that were
  achieved by means of our observing strategies and data analysis
  techniques. For \mbox{KOI-0410.01}, new transit data revealed a
  turn-over of its TTVs. We carried out an in-depth study of the
  system, that is identified in the NASA's Data Validation Report as
  false positive. Among others, we investigated a
  gravitationally-bound hierarchical triple star system, and a
  planet-star system. While the simultaneous transit fitting of ground
  and space-based data allowed for a planet solution, we could not
  fully reject the three-star scenario. New data, already scheduled in
  the upcoming 2018 observing season, will set tighter constraints on
  the nature of the system.

}  

   \maketitle
%
%________________________________________________________________

\section{Introduction}

Transit observations provide a wealth of information about alien
worlds. Beside the detection and characterization of exoplanets
\citep[e.g.][]{Seager2010}, once an exoplanet is detected by its
transits the variations of the observed mid-transit times can be used
to characterize the dynamical state of the system
\citep{Holman2005,Agol2005}. The timings of a transiting planet can
sometimes be used to derive constraints on the planetary physical and
orbital parameters in the case of multiple transiting planets
\citep{Holman2010}, to set constraints on the masses of the perturbing
bodies \citep{Ofir2014}, and to characterize the mass and orbit of a
non-transiting planet, with masses potentially as low as an Earth mass
\citep{Agol2005,Nesvorny2013,Barros2013arXiv,
  Kipping2014,Jontof-Hutter2015}. For faint stars, this is extremely
challenging to achieve by means of other techniques.

In the past three decades, non-Keplerian motions of exoplanets have
been regularly studied from the ground and space
\citep{Rasio1992,Malhotra1992,Peale1993,Wolszczan1994,Laughlin2001,Rivera2010,Holman2010,Lissauer2011a,Becker2015,Gillon2016}. Some
examples of ground-based transit timing variation (TTV) studies are
WASP-10b \citep{Maciejewski2011}, WASP-5b \citep{Fukui2011}, WASP-12b
\citep{Maciejewski2013}, and WASP-43b \citep{Jiang2016}. Accompanying
the observational growth, theoretical and numerical models were
developed to reproduce the timing shifts and represent the most
probable orbital configurations
\citep[e.g.,][]{Agol2005,Nesvorny2008,Lithwick2012,Deck2014}. There is
no doubt about the detection power of the TTV method: given the mass
of the host star, analyzing photometric observations we can sometimes
retrieve the orbital and physical properties of complete planetary
systems \citep{Carter2012}. However, the method requires sufficiently
long baseline, precise photometry and good phase coverage.

From ground-based studies, which have focused on TTVs of hot Jupiters,
there have already been some discrepant results \citep[see
  e.g. Qatar-1,][]{vonEssen2013,Mislis2015,Collins2017}, especially
when small-sized telescopes are involved and TTVs of low amplitude are
being measured \citep{vonEssen2016}. Also, many follow-up campaigns of
hot Jupiters could not significantly observe TTVs from the ground
\citep[see e.g.][for TrES-1, HAT-P-14b, WASP-28b, WASP-14b, and
  HAT-P-12b
  respectively]{Steffen2005,Fukui2016,Petrucci2015,Raetz2015,Mallonn2015}.
However, hot Jupiters tend to be isolated from companion planets
\citep{Steffen2012X} so it does not come as a surprise that these
studies have not resulted in convincing signals. It was with the
advent of space-based observatories that a new era in the TTV quest
started. In March 2009, NASA launched the Kepler space telescope
\citep{Borucki2010,Koch2010}. The main goal of the mission was to
detect Earth-sized planets in the so-called habitable zone, orbiting
around stars similar to our Sun. The wide field of view allowed
simultaneous and continuous monitoring of many thousands of stars for
about four years. Surprisingly, Kepler showed a bounty of planetary
systems with a much more compact configuration than our Solar System
\citep{Lissauer2014}. About 20\% of the known planetary systems
present either more than one planet or more than one star
\citep{Fabrycky2014}. Particularly, most multiple systems are formed
by at least two planets and about one third of these appear to be
close to mean motion resonant orbits
\citep[see][]{Lissauer2011b}. Thus, the long-term and highly precise
observations provided by Kepler have been the most successful data
source used to confirm and characterize planetary systems via
TTVs. Preceding a very long list, the first example of outstanding TTV
discoveries is \mbox{Kepler-9} \citep{Holman2010}. Since then, several
other planetary systems were confirmed, detected or even characterized
by means of TTV studies \citep[see
  e.g.][]{Hadden2014,Nesvorny2014}. Classic examples are
\mbox{Kepler-11} \citep{Lissauer2011a}, \mbox{Kepler-18}
\citep{Cochran2011}, \mbox{Kepler-19} \citep{Ballard2011},
\mbox{Kepler-23} and \mbox{Kepler-24} \citep{Ford2012},
\mbox{Kepler-25} to \mbox{Kepler-28} \citep{Steffen2012},
\mbox{Kepler-29} to \mbox{Kepler-32} \citep{Fabrycky2012}, and
\mbox{Kepler-36} \citep{Carter2012}. The list goes up to
\mbox{Kepler-87} \citep{Ofir2014} and continues with K2, Kepler's
second chance at collecting data that will allow us to investigate
planetary systems by means of TTVs \citep[see
  e.g.][]{Becker2015,Nespral2016,Jontof-Hutter2016,Hadden2017}. \cite{Mazeh2013}
analyzed the first twelve quarters of Kepler photometry and derived
the transit timings of 1960 Kepler objects of interest (KOIs). An
updated analysis of Kepler TTVs using the full long-cadence data set
can be found under \cite{Holczer2016}. The authors found that 130 KOIs
presented significant TTVs, either because their mid-transit times had
a large scatter, showed a periodic modulation, or presented a
parabola-like trend. Although $\sim$80 KOIs showed a clear sinusoidal
variation, for other several systems the periodic signal was too long
in comparison with the time span of Kepler data to cover one full TTV
cycle. As a consequence, no proper dynamical characterization could be
carried out.

To overcome this drawback and expand upon Kepler's heritage, in the
framework of a large collaboration we organized the Kepler Object of
Interest Network\footnote{koinet.astro.physik.uni-goettingen.de}
(KOINet). The main purpose of KOINet is the dynamical characterization
of selected KOIs showing TTVs. To date, the network is comprised of
numerous telescopes and is continuously evolving. KOINet's first light
took place in March, 2014. Here we show representative data obtained
during our first and second observing seasons that will highlight the
need for KOINet. Section~\ref{sec:KOINet} shows the basic working
structure of KOINet and the scientific milestones,
Section~\ref{sec:DATA} describes the observing strategy and the data
reduction process. Section~\ref{sec:RESULTS} makes special emphasis to
the fitting strategy of both ground and space-based data. In
Section~\ref{sec:MS} we show KOINet's achieved milestones, and we
finish with Section~\ref{sec:CONCLUSIONS}, where we present our
conclusions and a brief description of the future observing seasons of
KOINet.

\section{Kepler Object of Interest Network}
\label{sec:KOINet}

\subsection{Rationale}
\label{subsec:SC}

KOINet's unique characteristic is the use of already existing
telescopes, coordinated to work together towards a common goal. The
data collected by the network will provide three major contributions
to the understanding of the exoplanet population. First, deriving
planetary masses from transit timing observations for more planets
will populate the mass-radius diagram. The distribution of planetary
radii at a given planetary mass is surprisingly wide, revealing a
large spread in internal compositions \citep[see
  e.g.][]{Mordasini2012}. New mass and radius determinations will
provide new constraints for planet structure models. Furthermore,
longer transit monitoring will set tighter constraints for the
existence of non-transiting planets \citep{Barros2014}, providing a
broader and deeper view of the architecture of planetary
systems. Finally, a larger sample of well-constrained physical
parameters of planets and planetary systems will provide better
constraints for their formation and evolution
\citep{Lissauer2011b,Fang2012}.

KOINet is initially focusing its instrumental resources on 60 KOIs
that require additional data to complete a proper characterization or
validation by means of the TTV technique. Basic information on the
selected KOIs can be seen in the left part of
Table~\ref{tab:KOIs_params}. The KOI target list was built up based on
the work of \cite{Ford2012b}, \cite{Mazeh2013},
\cite{Xie2013,Xie2014}, \cite{Nesvorny2013}, \cite{Ofir2014}, and
\cite{Holczer2016}. The 60 KOIs were drawn from four groups, depending
on the scientific insights that further observations were expected to
provide.

For a pair of planets, an anti-correlation in the TTV signal is
expected to occur. This is the product of conservation of energy and
angular momentum and is stronger when the planetary pair is near
mean-motion resonance \citep[see e.g.][]{Holman2010,Carter2012,
  Lithwick2012}. The systems that present polynomial-shaped TTVs and
show anti-correlated TTV signals are given the highest priority,
independent of their status as valid planet candidates. In these
cases, any additional data points in their parabolic-shaped TTVs can
reveal a turn-over point, allowing a more accurate determination of
planetary masses. Further data will allow the analysis of the system's
dynamical characteristics. The systems that present anti-correlation
and a sinusoidal variation, but are poorly sampled, have second
priority (such as \mbox{KOI-0880.01/02}, a detailed analysis of the
system is in prep). In this case, more data points will allow us to
improve the dynamical analysis of these systems. Under third priority
fall the KOIs with very long TTV periodicity. Additional data might
shed some light into the constitution of these systems (for example,
\mbox{KOI-0525.01}, Section~\ref{sec:PA}). Finally, the lowest
priority is given to those systems that have been already
characterized, and the systems showing only one TTV signal (e.g.,
\mbox{KOI-0410.01}, Section~\ref{sec:TTVTO}). In the latter, under
specific conditions the perturber's mass and orbital period can be
constrained, confirming its planetary nature or ruling it out
\citep[e.g.,][]{Nesvorny2013, Nesvorny2014}.

\subsection{Observing time}

During the first two observing seasons (April-September, 2014 and
2015) an approximate total of 600 hours were collected for our
project, divided between 16 telescopes and 139 observing events.
Rather than following up all of the KOIs, we focused on the most
interesting ones from a dynamical point of view. Although here we
present a general overview of the data collected by KOINet and its
performance, we will focus in the analysis of individual KOIs in
upcoming publications.

\subsection{Basic characteristics of KOINet's telescopes}
\label{subsec:IC}

\begin{figure}[ht!]
  \centering
  \includegraphics[width=.5\textwidth]{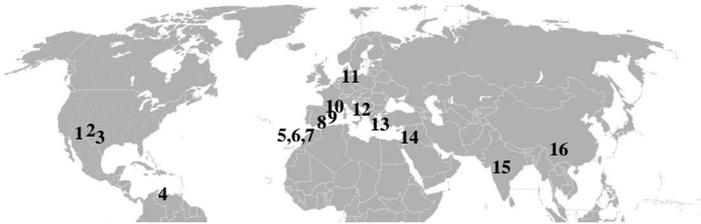}
  \caption{\label{fig:map} World's northern hemisphere, showing
    approximate locations of the observing sites that acquired data
    for KOINet during the 2014 and 2015 seasons. Numbers correspond to
    Table~\ref{tab:telescopes}.}
\end{figure}

Kepler planets and planet candidates showing TTVs generally present
two major disadvantages for ground-based follow-up observations. On
one hand, their host stars are relatively faint (K$_p\sim$12-16). On
the other hand, most of the KOIs reported to have large amplitude TTVs
produce shallow primary transits. To collect photometric data with the
necessary precision to detect shallow transits in an overall good
cadence, most of KOINet's telescopes have relatively large collecting
areas. This allows to collect data at a frequency of some seconds to a
few minutes. Another observational challenge comes with the transit
duration. For some of the KOIs the transit duration is longer than the
astronomical night, especially bearing in mind that the Kepler field
is best observable around the summer season, when the nights are
intrinsically shorter. In these cases full transit coverage can only
be obtained combining telescopes well separated in longitude. The
telescopes included in this collaboration are spread between America,
Europe and Asia, allowing almost 24 hours of continuous coverage. A
world map including the telescopes that collected data during 2014 and
2015 can be found in Figure~\ref{fig:map} and
Table~\ref{tab:telescopes}.

\begin{table}[ht!]
  \centering
  \caption{\label{tab:telescopes} Placement of the observatories that
    collected data during the 2014 and 2015 observing seasons.}
  \resizebox{0.45 \textwidth}{!}{
  \begin{tabular}{l l}
    \hline\hline
    1    &  Multiple Mirror Telescope Observatory (6.5m), United States of America \\
    2    &  Apache Point Observatory (3.5m), United States of America \\
    3    &  Monitoring Network of Telescopes (1.2m), United States of America \\
    4    &  Observatorio Astron\'omico Nacional del Llano del Hato (1m), Venezuela \\
    5    &  Nordic Optical Telescope (2.5m), Spain \\
    6    &  Liverpool Telescope (2m), Spain \\
    7    &  IAC80 telescope (0.8m), Instituto de Astrof\'isica de Canarias, Spain\\
    8    &  Calar Alto Observatory (1.25, 2.2, 3.5 m), Spain \\
    9    &  Planetary Transit Study Telescope (0.6m), Spain \\
    10   &  Joan Or\'o Telescope - The Montsec Astronomical Observatory (0.8m), Spain \\
    11   &  Oskar L\"uhning Telescope - Hamburger Sternwarte (1.2m), Germany \\
    12   &  Bologna Astronomical Observatory (1.52m), Italy \\
    13   &  Kryoneri Observatory - National Observatory of Athens (1.2m), Greece \\
    14   &  Wise Observatory - Tel-Aviv University (1m), Israel \\
    15   &  IUCAA Girawali Observatory (2m), India \\
    16   &  Yunnan Observatories (2.4m), China \\
    \hline
 \end{tabular}%
}
\end{table}

\begin{figure}[ht!]
  \centering
  \includegraphics[width=.5\textwidth]{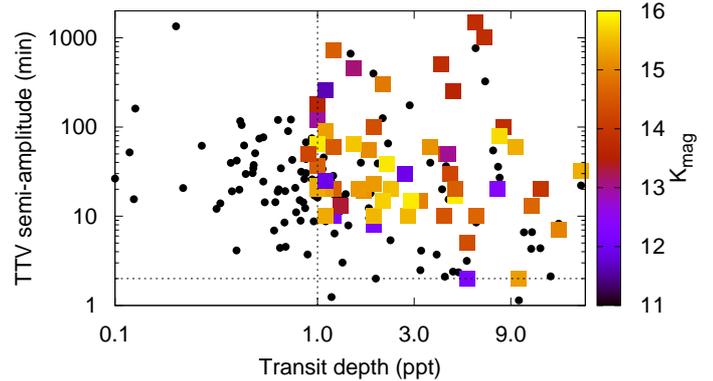}
  \caption{\label{fig:KOIs_TTVamp} Colored rectangles show TTV Kepler
    variability in minutes versus transit depth in ppt for the 60 KOIs
    that are included in KOINet. The squares are
    color-coded depending on the Kepler magnitude of the host
    star. Black circles show all the KOIs presenting TTVs with a
    Kepler variability larger than 1 minute. Vertical and horizontal
    dashed lines indicate the $\sim$1 ppt and 2 minutes limits for
    KOINet.}
\end{figure}

\noindent Considering these two fundamental limitations, to maximize
the use of KOINet data and boost transit detection we have included
the KOIs whose transit depth are larger than one part per thousand
(ppt) and which Kepler timing variability (this is, the variability
comprised within Kepler time span) is larger than two minutes (see
Figure~\ref{fig:KOIs_TTVamp}). Below these limits, the photometric
precision (and thus, the derived timing precision), and especially the
impact of correlated noise on photometric data \citep{Carter2009}
would play a fundamental role in the detection of transit
events. Next, we describe the primary characteristics of the
telescopes involved in this work.

\begin{itemize}
\item The Apache Point Observatory, located in New Mexico, United
  States of America, hosts the Astrophysical Research Consortium 3.5
  meter telescope, henceforth \mbox{ARC 3.5m}. The data were collected
  using Agile \citep{Mukadam2011}. Concerning the data presented in
  this work, the \mbox{ARC 3.5m} observed one transit of
  \mbox{KOI-0525.01}, our lower-limit KOI for transit
  depth. Nonetheless, during the first observing seasons we have
  collected a substantial amount of data that will be presented in
  future work.
\item The Nordic Optical Telescope (henceforth \mbox{NOT 2.5m}) is
  located at the observatory ``Roque de los Muchachos'' in La Palma,
  Spain, and belongs to the Nordic Optical Telescope Scientific
  Association, governed and funded by Scandinavian countries. In this
  work we present observations of \mbox{KOI-0760.01} and
  \mbox{KOI-0410.01}.
\item The 2.2m Calar Alto telescope is located in Almer\'ia, Spain
  (henceforth \mbox{CAHA 2.2m}). We observed \mbox{KOI-0410.01} using
  the Calar Alto Faint Object Spectrograph in its photometric mode.
\item The IAC80 telescope (henceforth, \mbox{IAC 0.8m}) is located at
  the Observatorio del Teide, in the Canary Islands, Spain. We
  observed half a transit of \mbox{KOI-0902.01} for about 7 hours.
\item The Lijiang 2.4m telescope (henceforth \mbox{YO 2.4m}) is
  located at the Yunnan Observatories in Kunming, China. In
  this work we present observations of \mbox{KOI-0410.01}.
\end{itemize} 

\subsection{Maximizing the use of KOINet's time}

\subsubsection{Assigning telescopes to KOIs}
\label{sec:tel_time}

In order to effectively distribute the available telescope time and
maximize our chances to detect transit events, three main
characteristics have to be considered: the apparent magnitude of the
host star, the available collecting area given by the size of the
primary mirror, and the amplitude and scatter of Kepler TTVs. With the
main goal to connect the KOIs to the most suitable telescopes, we
proceed as follows. First, we estimate the exposure time,
$\mathrm{E_t}$, for each host star and telescope. The latter is
computed to achieve a given signal-to-noise ratio (SNR), so that
\mbox{SNR = 1/T$_{\mathrm{depth}}$} is satisfied. In this case,
T$_{\mathrm{depth}}$ corresponds to the transit depth in percentage,
which is taken from the NASA Exoplanet
Archive\footnote{https://exoplanetarchive.ipac.caltech.edu}. Besides
the desired signal-to-noise ratio, the calculation of $\mathrm{E_t}$
is carried out considering parameters such as the mean seeing of the
site, the brightness of the star, the size of the primary mirror,
typical sky brightness of the observatories, the phase of the Moon,
and the altitude of the star during the predicted observing
windows. Once the exposure times are computed, the derived values are
verified and subsequently confirmed by each telescope leader.

Off-transit data have a considerable impact in the determination of
the orbital and physical parameters of any transiting system. In the
case of ground-based observations, off-transit data are critical to
remove systematic effects related to changes in airmass,
color-dependent extinction, and poor guiding and flatfielding
\citep[see e.g.,][]{Southworth2009,vonEssen2016}. Henceforth, to
determine the number of data points per transit, $N$, we use the
estimated exposure time and the known transit duration,
$\mathrm{T_{dur}}$, incremented by two hours. This increment accounts
for 1 hour of off-transit data before and after transit begins and
ends, respectively. Then, the number of data points per transit is
simply estimated as \mbox{$N = \mathrm{(T_{dur} + 2~ hs)/(E_t +
    ROT)}$}. Here, ROT corresponds to the readout time of
charge-coupled devices used to carry out the observations. To compute
the timing precision, $\mathrm{\sigma_T}$, we use a variant of the
formalism provided by \cite{Ford2007}:

\begin{equation}
\mathrm{\sigma_T = \frac{Phot_P \times T_{dur}}{N^{1/2} \times T_{depth}}},\
\end{equation}

\noindent where $\mathrm{Phot_P}$ is the photometric precision in
percentage that a given telescope can achieve while observing a
\mbox{14-15 $\mathrm{K_p}$ star}. This value was requested to the
members of KOINet immediately after they joined the network. Comparing
the estimated timing precision with the semi-amplitude of Kepler TTVs
\mbox{(A$\mathrm{_{TTVs}> 3\sigma_T}$)} yields erroneous results,
especially if the TTVs are intrinsically large. For example, an
estimated timing precision of one hour satisfies the above condition
for a TTV semi-amplitude of 3 hours. However, when ground-based
photometry is being analyzed, a timing precision of one hour would be
equal to a non-detection. Therefore, to assign a KOI to a telescope
three aspects are simultaneously considered: the transit depth
\mbox{($\mathrm{T_{depth}>\mathrm{Phot_P}}$)}, the amplitude
of Kepler TTVs \mbox{(A$\mathrm{_{TTVs}> 3\sigma_T}$)}, and the
natural scatter of Kepler TTVs \mbox{($\mathrm{2\sigma_{TTVs} >
    \sigma_T}$)}.

\subsubsection{Prescription for optimum reference stars}

Differential photometry highlights the variability of one star (the
so-called target star) relative to another one (the reference star)
which ideally should not vary in time. Thus, the selection of
reference stars can limit the precision of photometric data
\citep{Young1991,Howell2006}. The true constancy of reference stars is
given by how much they intrinsically vary, subject to the precision
that a given optical setup can achieve. Analyzing the flux
measurements along the 17 quarters of all the stars within a radius of
5 arcmin relative to KOINet's KOIs, we selected stars that showed a
constant flux behavior in time and had a comparable brightness to the
given KOI \citep{Howell2006}. In this way, we provide to the observer
the location of the most photometrically well-behaved reference stars,
minimizing the noise budget right from the beginning. Particularly, we
have identified between 2 to 5 reference stars per field of view, and
their location on sky are provided to the observers through KOINet's
web interface.

\subsection{Predictions computed from Kepler timings}

\begin{figure*}[ht!]
  \centering
  \includegraphics[width=.47\textwidth]{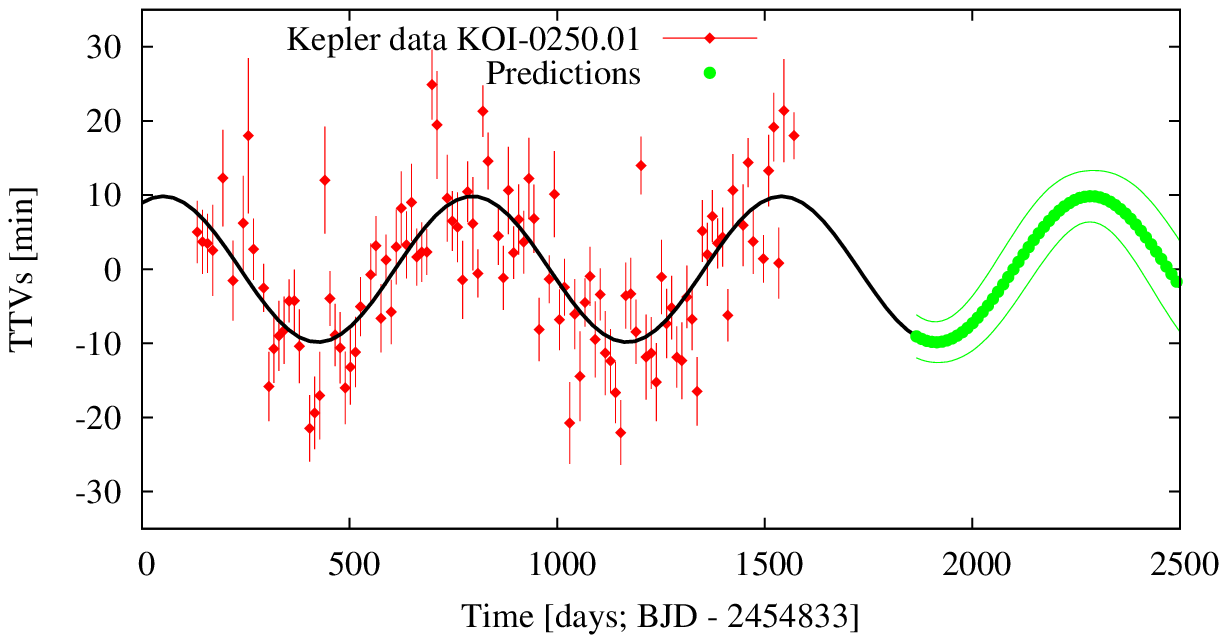}
  \includegraphics[width=.47\textwidth]{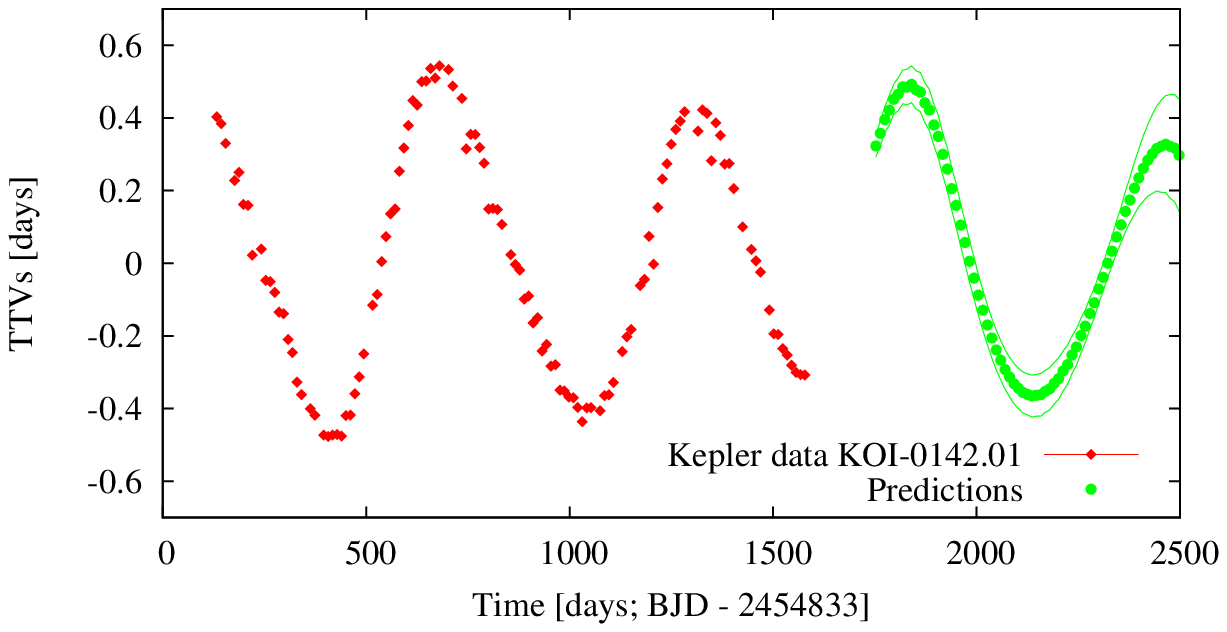}

  \includegraphics[width=.47\textwidth]{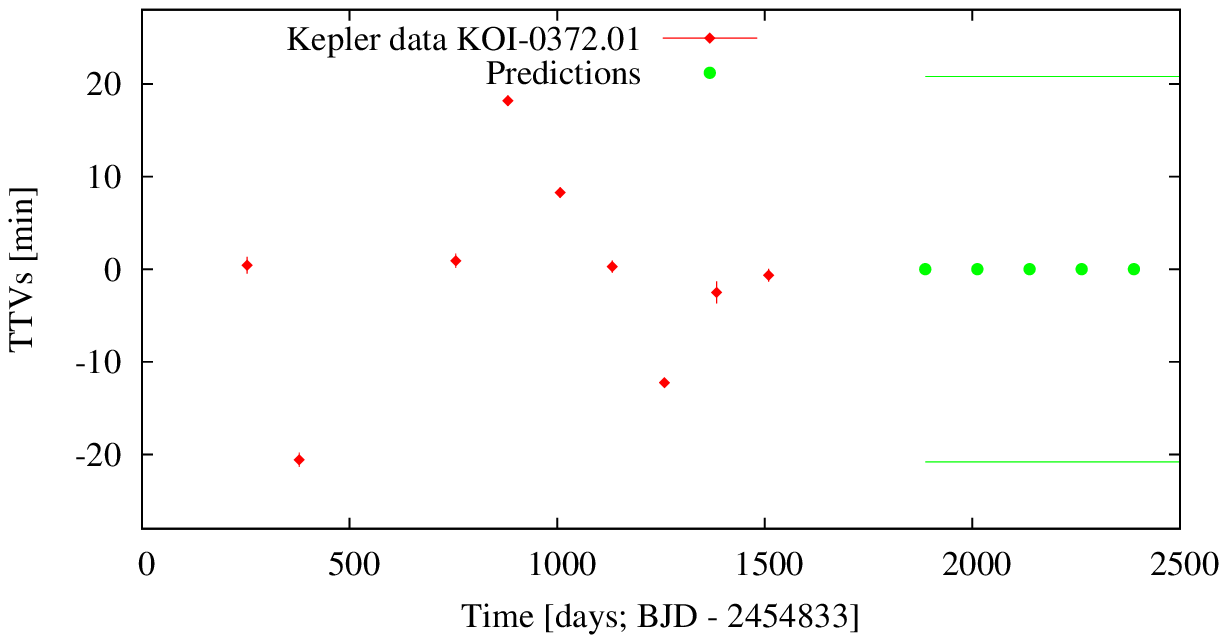}
  \includegraphics[width=.47\textwidth]{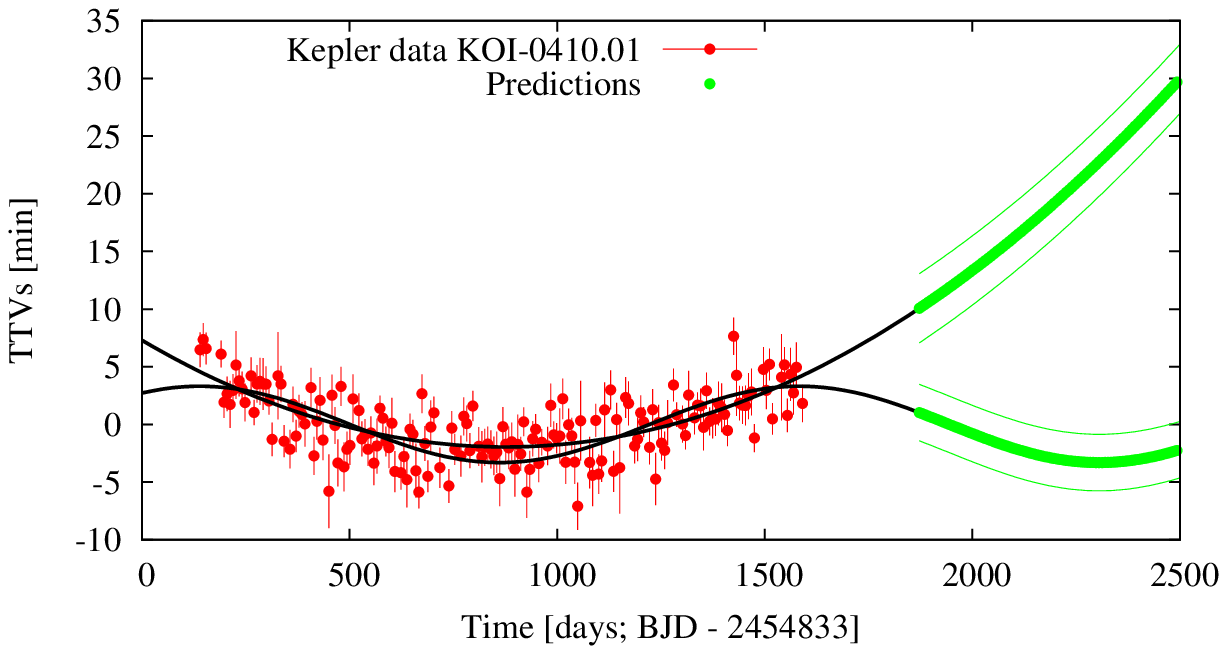}
  \caption{\label{fig:cases_TTVs} {\it From left to right and top to
      bottom:} sinusoidal, dynamic, chaotic, and parabolic/sinusoidal
    classification of the TTVs. Note that TTVs for \mbox{KOI-0142.01}
    are given in days, rather than minutes.}
\end{figure*}

Using the mid-transit times obtained from Kepler 17 quarters, we
computed TTVs subtracting from them an averaged (constant) period, and
classified the KOIs depending on the shape of their TTVs. A full
description of the fitting process of Kepler transit light curves, the
derived values, and their associated errors, can be found in
Section~\ref{sec:PTF1}. For now, Figure~\ref{fig:cases_TTVs} shows our
four target groups. The simplest case, in which the TTVs
follow a sinusoidal shape, is shown on the top left panel of the
Figure. To estimate the predictions for our ground-based follow-up we
fitted to Kepler mid-transit times a linear plus a sinusoidal term:
\begin{equation}
  TTV(E) = T_0(E=0) + P_C\times E + A\times \sin[2\pi(\nu~E + \phi)].\
  \label{eq:SINUS}
\end{equation}
\noindent In this case, E corresponds to the transit epoch,
$\mathrm{T_0(E=0)}$ to a reference mid-transit time, $\mathrm{P_C}$ is
the orbital (constant) period, A the semi-amplitude of the TTVs, and
$\nu$ and $\phi$ the frequency and phase of the TTVs,
respectively. The derived predictions are shown in
Figure~\ref{fig:cases_TTVs} in green points, while Kepler data is
plotted in red and the shape of the predictions, including Kepler
time, is shown in continuous black line. Since all Kepler mid-times
show some scatter, we also estimated errors in the predictions taking
this noise into consideration. To increase the chance of transit
detection, the magnitudes of the errors in the predictions are
provided to the observers, along with a warning. The second TTV
scenario is shown in the top right panel of
Figure~\ref{fig:cases_TTVs}. In this case the available data and the
systems themselves allow a more refined dynamical analysis of the TTVs
by means of n-body simulations and/or simultaneous transit fitting
\citep[see e.g.,][]{Agol2005,Nesvorny2013,Nesvorny2014}, from which
the predictions are computed. Due to their complexity, a detailed
description of the computation of these TTVs is beyond the scope of
this paper, and will be given individually in future publications. The
third case is shown in the bottom left panel of
Figure~\ref{fig:cases_TTVs}. Here, the number of available Kepler
transits is not sufficient to carry out a dynamical analysis, and the
TTVs don't follow any shape that could give us a hint of when could
the upcoming transits occur. Thus, to determine the predictions we fit
to Kepler mid-times a linear trend only (i.e., assuming constant
period), and use as errors for the predictions the semi-amplitude of
the TTVs. The last case exemplifies the need for a ground-based
campaign taking place immediately after Keplers follow-up. This case,
displayed in the bottom right panel of Figure~\ref{fig:cases_TTVs},
shows an incomplete coverage of the TTV periodicity. From photometry
only we cannot assess if the cause for TTVs is planetary in nature, is
gravitationally bound to the system (e.g., TTVs following a sinusoidal
shape), or some completely different scenario, like TTVs caused by a
blended eclipsing binary (TTVs showing a parabolic shape). In this
case, we produce two kind of predictions: sine TTVs, from where the
predictions are computed as described in Eq.~\ref{eq:SINUS}, and
parabolic TTVs:
\begin{equation}
  TTV(E) = T_0(E=0) + P_C\times E + a\times E^2 + b\times E + c.\
  \label{eq:PAR}
\end{equation}
\noindent where $a$, $b$, and $c$ are the fitting coefficients of the
parabola. Although these are the two scenarios most likely to occur,
the mid-times could also show a different trend. Therefore, until we
can disentangle which trend is the one that the system follows, we
provide to the observers both predictions and ask them to observe both
of them, and extend the observing time as much as they can.

\section{Observations and data reduction}
\label{sec:DATA}

\subsection{Basic observing setup}

In order to ensure observations as homogeneous as possible, observers
are asked to carry them out in a specific way. To begin with, our
observations cover a range of airmass and so are subject to
differential extinction effects between the target and comparison
stars. To minimize color-dependent systematic effects observers used
intermediate (Cousins R) or narrow-band (gunn r) filters, depending on
the brightness of the target stars and filter availability. The use of
R-band filters also reduces light curve variations from starspots and
limb-darkening effects, and they circumvent the large telluric
contamination around the I-band. Furthermore, all observers provide
regular calibrations (bias flatfield frames and darks, if needed), and
are asked to observe with the telescope slightly defocused to minimize
the noise in the photometry \citep{Kjeldsen1992,Southworth2009}. Once
the observations are performed, they are collected and reduced in an
homogeneous way.

\subsection{DIP$^2$OL}
\label{sec:photom}

KOINet data are reduced and analyzed by means of the {\it Differential
  Photometry Pipelines for Optimum Lightcurves}, DIP$^2$OL. The
pipeline is divided in two parts. The first one is based in IRAF's
command language. It requires only one reference frame to do aperture
photometry. The pipeline carries out normal calibration sequences
(bias and dark subtraction and flatfield division, depending on
availability) using IRAF task ccdproc. In the particular case of
KOINet data, acquired calibrations are always a set of bias and
flatfields, taken either at the beginning or end of each observing
night. Subject to availability, we correct the science frames of a
given observing night with their corresponding calibrations only. In
general, we do not take dark frames due to short exposures and cooled,
temperature stable CCDs. The reduction continues with cosmic rays
rejection (IRAF's cosmicrays) and alignment of the science frames
(imalign). Afterwards, reference stars within the field are chosen
following specific criteria \citep[for example, that the brightness of
  the reference stars have to be similar to the brightness of the
  target star to maximize the signal-to-noise ratio of the
  differential light curves,][]{Howell2006} and photometric fluxes and
errors are measured over the target star and the reference stars as a
function of 10 different aperture radii and 3 different sky rings. The
annulus and the initial width of the sky ring are set by the user,
since they depend on the crowding of the fields. The apertures are
non-uniformly distributed between 0.5 and \mbox{5$\times\hat{S}$},
with more density between 1 and \mbox{2$\times\hat{S}$}. Here,
$\hat{S}$ corresponds to the averaged seeing of the images, computed
from the full-width at half maximum of all the chosen stars in the
field. This, in turn, sets a limit to the lowest possible value for
the annulus. To perform a posterior detrending of the photometric
data, in addition to $\hat{S}$, the pipeline computes the airmass
corresponding to the center of the field of view, the (x,y) centroid
positions of all the measured stars, three sky values originally used
to compute the integrated fluxes (one per sky ring), and the
integrated counts of the master flat and master dark over the (x,y)
values per frame and per aperture.

The second part of DIP$^2$OL is python-based. The routine starts by
producing \mbox{N+1} light curves from the N reference fluxes
previously computed by IRAF, one with the summed flux of all the N
comparison stars and N versions with all the reference stars except
one. If one of the reference stars is photometrically unstable, the
residual light curve corresponding to the unweighted sum of the fluxes
of all the reference stars minus this one will show up by giving the
lowest standard deviation, when compared to the remaining N
residuals. Therefore, this star is removed from the sample. The
process of selection and rejection is repeated until the combination
of the current available reference stars gives the lowest scatter in
the photometry. Since a priori we don't know if primary transits are
actually observed within a given predicted window, residuals are
computed by dividing the differential fluxes by a spline function. The
pipeline repeats this process through all measured apertures and sky
rings, and finds the combination of reference stars, aperture and sky
ring that minimizes the standard deviation of the differential light
curves \citep[see e.g.,][]{Ofir2014}. Finally, the code outputs the
time in Julian dates shifted to the center of the exposure, the
differential fluxes, photometric error bars which magnitudes have been
scaled to match the standard deviation of the residuals, (x,y) centroid
positions, flat counts that were integrated within the final aperture
around the given centroids, sky fluxes corresponding to the chosen sky
ring, and seeing and airmass values. These quantities will be used in
a following step to compute the ground-based detected mid-transit
times.

\section{Data modelling and fitting strategies}
\label{sec:RESULTS}

\subsection{Primary transit fitting of Kepler data}
\label{sec:PTF1}

\begin{figure}[ht!]
  \centering
  \includegraphics[width=.45\textwidth]{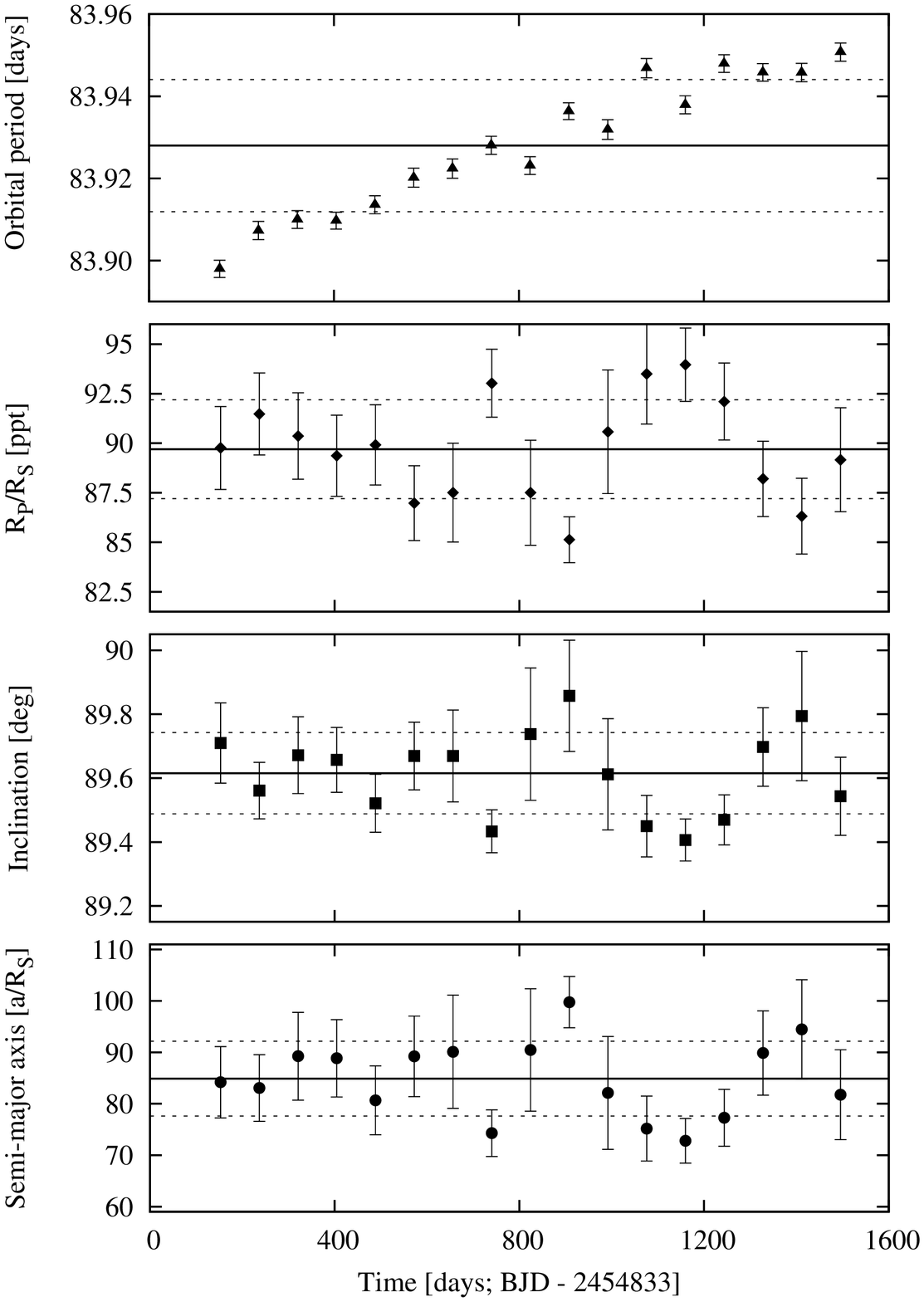}
  \caption{\label{fig:ParsVar} Time-dependent change of the transit
    parameters of \mbox{KOI-0902.01}. From top to bottom the orbital
    period in days in triangles, the planet-to-star radius ratio,
    $\mathrm{R_P/R_S}$ in diamonds, the orbital inclination in
    squares, and the semi-major axis in stellar radii,
    $\mathrm{a/R_S}$. Horizontal continuous and dashed lines show mean
    and standard deviations of the system parameters,
    respectively. Individual errors are given at 1-$\sigma$ level.}
\end{figure}

One of the key ingredients for the success of our ground-based TTV
follow-up is the prior knowledge, with a good degree of accuracy, of
the orbital and physical parameters of the systems. To take full
advantage of Kepler data in our work, we re-computed the orbital and
physical parameters of the 60 KOIs that are included in KOINet's
follow-up. A quick view into the Data Validation Reports suggested us
that the procedures performed over KOIs without TTVs was not optimum
for KOIs showing TTVs. Thus, we did not use the transit parameters
reported by the NASA Exoplanet Archive. Rather than computing
time-expensive photo-dynamical solutions over the 60 KOIs \citep[see
  e.g.,][]{Barros2015}, to minimize the impact of the TTVs in the
computation of the transit parameters we fitted two consequent transit
light curves simultaneously with a \cite{MandelAgol2002} transit
model, making use of their \texttt{occultquad}
routine\footnote{\url{http://www.astro.washington.edu/users/agol}}. From
the transit light curve we can determine the following parameters: the
orbital period, Per, the mid-transit time, $\mathrm{T_0}$, the
planet-to-star radius ratio, $\mathrm{R_p/R_s}$, the semi-major axis
in stellar radius, $\mathrm{a/R_s}$, and the orbital inclination, i,
in degrees. For all the KOIs we assumed circular orbits. Furthermore,
we assumed a quadratic limb-darkening law with fixed limb darkening
coefficients, $\mathrm{u_1}$ and $\mathrm{u_2}$. For the Kepler data
we used the limb-darkening values specified in \cite{Claret2013},
choosing as fundamental stellar parameters, effective temperature,
metallicity and surface gravity, the values listed in the NASA
Exoplanet Archive. Simultaneously to the transit model we fitted a
time-dependent second-order polynomial to account for out-of-transit
variability. To determine reliable errors for the fitted parameters,
we explored the parameter space by sampling from the
posterior-probability distribution using a Markov-chain Monte-Carlo
(MCMC) approach. Our MCMC calculations make extensive use of routines
of
\texttt{PyAstronomy}\footnote{\url{http://www.hs.uni-hamburg.de/DE/Ins/Per/Czesla/}\\ \url{PyA/PyA/index.html}},
a collection of Python build-in functions that provide an interface
for fitting and sampling algorithms implemented in the PyMC
\citep{Patil2010} and SciPy \citep{Jones2001} packages. We refer the
reader to their detailed online
documentation\footnote{\url{http://pymc-devs.github.io/pymc/}}. For
the computation of the best-fit parameters we iterated \mbox{80\ 000}
times per consecutive transits, and discarded a conservative first
20\%. As starting values for the parameters we used the ones specified
in the NASA Exoplanet Archive. To set reasonable limits for MCMC's
uniform probability distributions, we chose \mbox{R$_P$/R$_S$ $\pm$
  0.1}, \mbox{T$\mathrm{_0}$ $\pm$ T$_{\mathrm{Dur}}$/3}, and a
considerable fraction of the orbital period, depending on the
amplitude of Kepler TTVs. These values are relative to the values
determined by the Kepler team. The semi-major axis and the inclination
are correlated through the impact parameter, \mbox{$\mathrm{a/R_S
    cos(i)}$}. Thus, rather than using uniform distributions for these
parameters we used Gaussian priors with the mean and the standard
deviation equal to the values found in the NASA Exoplanet Archive and
three times their errors, respectively. To compute the transit
parameters we analyzed Kepler long cadence transit data. To minimize
the impact of the sampling rate on the determination of the transit
parameters \citep[see e.g.,][]{Kipping2010}, during each instance of
primary transit fitting we used a transit model calculated from a
finer time scale and then averaged on the Kepler timing points. In
particular, 30 equally spaced points were calculated and averaged to
one data point. The modeling of all consecutive transits results in a
parameter distribution for the semi-major axis, the inclination, the
orbital period and the planet-to-star radius ratio. We used their mean
values and standard deviations to limit the ground-based data fitting
(Section~\ref{sec:PTF2}). All the orbital and physical parameters
computed for the 60 KOIs are summarized in the right part of
Table~\ref{tab:KOIs_params}. Errors are at the 1-$\sigma$ level. It is
worth to mention that the transit parameters presented in the table
provide us with an excellent transit template to be used to fit
ground-based data. It is not our intention to improve any of the
parameters by means of this simple analysis. A more detailed approach,
such as photo-dynamical fitting might be required \citep[see
  e.g.][]{Barros2015}, specially with large-amplitude TTVs such as
Kepler-9 \citep[KOI-0377.01/02,][]{Holman2010,Ofir2014}. As an
illustrative example, Figure~\ref{fig:ParsVar} shows how the transit
parameters change as a function of time, evidencing their mutual
correlations and the rate and amplitude at which they change. As
expected, for the values in the figure the Pearson's correlation
coefficient between the semi-major axis and the inclination is
\mbox{r$_{a/R_s,i}$ = 0.96}, while these two reveal a strong
anti-correlation with the planet-to-star radius ratio
(\mbox{r$_{R_p/R_s,i}$ = -0.91}, and \mbox{r$_{R_p/R_s,a/R_s}$ =
  -0.93}).

\begin{table*}[ht!]
    \caption{\label{tab:KOIs_params} {\it Left:} From left to right
      the KOI number, the right ascention, $\alpha$, and the
      declination, $\delta$, in degrees (J2000.0) and the Kepler
      magnitude, K$_p$. The values have been taken from the NASA
      Exoplanet Archive. {\it Right:} Best-fit orbital parameters
      obtained fitting all available primary transits from quarter 1
      to quarter 17 as described in this section. From left to right
      the semi-major axis in stellar radii, $\mathrm{a/R_S}$, the
      inclination in degrees, i, the planet-to-star radius ratio,
      $\mathrm{R_P/R_S}$, and the orbital period in days, Per. The
      last column, O14-15, corresponds to the number of observations
      collected during 2014 and 2015.}  \centering
  \begin{tabular}{c c c c | c c c c c}
    \hline \hline
    KOI   &   $\alpha$ (J2000) & $\delta$ (J2000) &   K$_p$  &   $\mathrm{a/R_S}$   &       i      &   $\mathrm{R_P/R_S}$   &  Per    & O14-15 \\
    Nr.   & ($^{\circ}$)         & ($^{\circ}$)      &          &             &  ($^{\circ}$)  &              & (days)  &                       \\
    \hline
0094.01 & 297.333069 & 41.891121 & 12.205 	& 27.27 $\pm$ 0.03 & 89.997 $\pm$ 0.001 & 0.0691 $\pm$ 0.0001 & 22.34285 $\pm$  0.00078  & -\\
0094.03 & 297.333069 & 41.891121 & 12.205 	& 50.5 $\pm$ 0.2 & 89.93 $\pm$ 0.01 & 0.0411 $\pm$ 0.0003 & 54.3198 $\pm$  0.0018 & - \\
0142.01 & 291.148071 & 40.669399 & 13.113 	& 16.9 $\pm$ 0.9 & 87.4 $\pm$ 0.3 & 0.038 $\pm$ 0.001  & 10.947 $\pm$  0.036 & 6 \\
0250.01 & 284.940979 & 46.566540 & 15.473 	& 32 $\pm$ 2 & 89.29 $\pm$ 0.07 & 0.051 $\pm$ 0.002  & 12.2827 $\pm$  0.0044 & 2 \\
0250.02 & 284.940979 & 46.566540 & 15.473 	& 54 $\pm$ 6 & 89.3 $\pm$ 0.2 & 0.047 $\pm$ 0.005  & 17.2509 $\pm$  0.0097 & 1 \\
0315.01 & 297.271881 & 43.333309 & 12.968 	& 59 $\pm$ 5 & 89.6 $\pm$ 0.2 & 0.029 $\pm$ 0.001  & 35.5812 $\pm$  0.0087 & 4 \\
0318.01 & 288.153992 & 44.068821 & 12.211 	& 29.1 $\pm$ 0.2 & 89.9 $\pm$ 0.2 & 0.033 $\pm$ 0.003  & 38.5846 $\pm$  0.0049 & - \\
0345.01 & 286.524811 & 48.683601 & 13.340 	& 45 $\pm$ 3 & 89.4 $\pm$ 0.2 & 0.0335 $\pm$ 0.0009 & 29.8851 $\pm$  0.0031 & - \\
0351.01 & 284.433502 & 49.305161 & 13.804 	& 186.2 $\pm$ 0.1 & 89.970 $\pm$ 0.001 & 0.0852 $\pm$ 0.0001 & 331.616 $\pm$  0.025 & 1 \\
0351.02 & 284.433502 & 49.305161 & 13.804 	& 141 $\pm$ 1 & 90.001 $\pm$ 0.001 & 0.0601 $\pm$ 0.0008 & 210.79 $\pm$  0.41 & 1 \\
0372.01 & 299.122437 & 41.866760 & 12.391 	& 112 $\pm$ 1 & 89.98 $\pm$ 0.08 & 0.0816 $\pm$ 0.0009 & 125.6287 $\pm$  0.0073 & - \\
0377.01 & 285.573975 & 38.400902 & 13.803 	& 33 $\pm$ 2 & 89.1 $\pm$ 0.2 & 0.078 $\pm$ 0.001  & 19.245 $\pm$  0.023 & 12 \\
0377.02 & 285.573975 & 38.400902 & 13.803 	& 55 $\pm$ 6 & 89.3 $\pm$ 0.2 & 0.076 $\pm$ 0.003  & 38.95 $\pm$  0.11 & 2 \\
0410.01 & 292.248016 & 40.696049 & 14.454 	& 33 $\pm$ 6 & 89.0 $\pm$ 0.9 & 0.065 $\pm$ 0.007  &  7.2165 $\pm$  0.0018 & 6 \\
0448.02 & 297.070160 & 40.868790 & 14.902 	& 45 $\pm$ 10 & 88.9 $\pm$ 0.6 & 0.05 $\pm$ 0.01 & 43.587 $\pm$  0.022 & 9 \\
0456.01 & 287.773560 & 42.869282 & 14.619 	& 20.4 $\pm$ 0.7 & 88.35 $\pm$ 0.07 & 0.034 $\pm$ 0.001 & 13.699 $\pm$  0.012 & 3 \\
0464.01 & 293.747101 & 45.107220 & 14.361 	& 75.0 $\pm$ 0.3 & 89.95 $\pm$ 0.01 & 0.0677 $\pm$ 0.0008 & 58.3619 $\pm$  0.0023 & - \\
0523.01 & 286.047119 & 45.053211 & 15.000 	& 45 $\pm$ 5 & 88.9 $\pm$ 0.2 & 0.063 $\pm$ 0.003 & 49.4112 $\pm$  0.0082 & 1 \\
0525.01 & 300.907776 & 45.457870 & 14.539 	& 20 $\pm$ 2 & 87.3 $\pm$ 0.3 & 0.05 $\pm$ 0.01 & 11.5300 $\pm$  0.0093 & 4 \\
0528.02 & 287.101105 & 46.896481 & 14.598 	& 102 $\pm$ 9 & 89.6 $\pm$ 0.1 & 0.031 $\pm$ 0.002 & 96.676 $\pm$  0.010 & - \\
0620.01 & 296.479767 & 49.937679 & 14.669 	& 62.7 $\pm$ 0.4 & 89.90 $\pm$ 0.02 & 0.074 $\pm$ 0.001 & 45.1552 $\pm$  0.0028 & 1 \\
0620.02 & 296.479767 & 49.937679 & 14.669 	& 127.2 $\pm$ 0.6 & 89.98 $\pm$ 0.01 & 0.1017 $\pm$ 0.0009 & 130.1783 $\pm$  0.0058 & - \\
0638.01 & 295.559418 & 40.236271 & 13.595 	& 36.1 $\pm$ 0.3 & 89.65 $\pm$ 0.06 & 0.032 $\pm$ 0.001 & 23.6415 $\pm$  0.0069 & 2 \\
0738.01 & 298.348328 & 47.491230 & 15.282 	& 27 $\pm$ 4 & 88.79 $\pm$ 0.07 & 0.037 $\pm$ 0.003 & 10.338 $\pm$  0.015 & 4 \\
0738.02 & 298.348328 & 47.491230 & 15.282 	& 24 $\pm$ 2 & 88.33 $\pm$ 0.05 & 0.034 $\pm$ 0.005 & 13.286 $\pm$  0.019 & - \\
0757.02 & 286.999481 & 48.375790 & 15.841 	& 68 $\pm$ 2 & 89.73 $\pm$ 0.07 & 0.046 $\pm$ 0.003 & 41.196 $\pm$  0.011 & - \\
0759.01 & 285.718536 & 48.504849 & 15.082 	& 37 $\pm$ 4 & 88.8 $\pm$ 0.3 & 0.044 $\pm$ 0.003 & 32.628 $\pm$  0.017 & 3 \\
0760.01 & 292.167053 & 48.727589 & 15.263 	& 12.2 $\pm$ 0.4 & 86.0 $\pm$ 0.2 & 0.106 $\pm$ 0.003 &  4.9592 $\pm$  0.0012 & 7 \\
0806.01 & 285.283630 & 38.947281 & 15.403 	& 124 $\pm$ 7 & 89.84 $\pm$ 0.09 & 0.099 $\pm$ 0.001 & 143.200 $\pm$  0.059 & 3 \\
0806.02 & 285.283630 & 38.947281 & 15.403 	& 75 $\pm$ 3 & 89.9 $\pm$ 0.1 & 0.136 $\pm$ 0.003 & 60.3258 $\pm$  0.0062 & 4 \\
0829.03 & 290.461761 & 40.562462 & 15.386 	& 37 $\pm$ 3 & 88.7 $\pm$ 0.1 & 0.033 $\pm$ 0.003 & 38.557 $\pm$  0.024 & - \\
0841.01 & 292.236755 & 41.085880 & 15.855 	& 31.3 $\pm$ 0.9 & 89.21 $\pm$ 0.07 & 0.054 $\pm$ 0.004 & 15.334 $\pm$  0.011 & 1 \\
0841.02 & 292.236755 & 41.085880 & 15.855 	& 39 $\pm$ 5 & 88.9 $\pm$ 0.3 & 0.08 $\pm$ 0.01 & 31.3304 $\pm$  0.0077 & 3 \\
0854.01 & 289.508484 & 41.812119 & 15.849 	& 89 $\pm$ 5 & 89.75 $\pm$ 0.05 & 0.041 $\pm$ 0.002 & 56.052 $\pm$  0.021 & 1 \\
0869.02 & 291.638977 & 42.436321 & 15.599 	& 57 $\pm$ 2 & 89.64 $\pm$ 0.08 & 0.037 $\pm$ 0.001 & 36.277 $\pm$  0.027 & 1 \\
0880.01 & 292.873383 & 42.966141 & 15.158 	& 36 $\pm$ 5 & 88.7 $\pm$ 0.4 & 0.045 $\pm$ 0.006 & 26.4435 $\pm$  0.0097 & 1 \\
0880.02 & 292.873383 & 42.966141 & 15.158 	& 51 $\pm$ 6 & 89.3 $\pm$ 0.2 & 0.061 $\pm$ 0.002 & 51.537 $\pm$  0.021 & 7 \\
0886.01 & 294.773926 & 43.056301 & 15.847 	&  8.9 $\pm$ 0.4 & 83.7 $\pm$ 0.1 & 0.07 $\pm$ 0.02 &  8.009 $\pm$  0.015 & 1 \\
0902.01 & 287.852386 & 43.897991 & 15.754 	& 85 $\pm$ 7 & 89.6 $\pm$ 0.1 & 0.089 $\pm$ 0.002 & 83.927 $\pm$  0.016 & 7 \\
0918.01 & 283.977509 & 44.811562 & 15.011 	& 51.6 $\pm$ 0.2 & 89.93 $\pm$ 0.02 & 0.116 $\pm$ 0.002 & 39.6432 $\pm$  0.0016 & 3 \\
0935.01 & 294.023010 & 45.853081 & 15.237 	& 30.9 $\pm$ 0.5 & 89.5 $\pm$ 0.1 & 0.042 $\pm$ 0.001 & 20.860 $\pm$  0.011 & 1 \\
0935.02 & 294.023010 & 45.853081 & 15.237 	& 40 $\pm$ 3 & 89.0 $\pm$ 0.2 & 0.042 $\pm$ 0.002 & 42.6334 $\pm$  0.0081 & - \\
0935.03 & 294.023010 & 45.853081 & 15.237 	& 52 $\pm$ 7 & 89.1 $\pm$ 0.3 & 0.034 $\pm$ 0.002 & 87.647 $\pm$  0.019 & - \\
0984.01 & 291.048798 & 36.839882 & 11.631 	& 20.6 $\pm$ 0.7 & 88.8 $\pm$ 0.1 & 0.030 $\pm$ 0.001 &  4.2888 $\pm$  0.0031 & 8 \\
1199.01 & 293.743927 & 38.939281 & 14.887 	& 72 $\pm$ 2 & 89.77 $\pm$ 0.08 & 0.030 $\pm$ 0.002 & 53.526 $\pm$  0.021 & - \\
1271.01 & 294.265503 & 44.794300 & 13.632 	& 105 $\pm$ 3 & 89.64 $\pm$ 0.02 & 0.0693 $\pm$ 0.0005 & 161.98 $\pm$  0.16 & 4 \\
1353.01 & 297.465332 & 42.882839 & 13.956 	& 112 $\pm$ 3 & 89.85 $\pm$ 0.07 & 0.105 $\pm$ 0.001 & 125.8648 $\pm$  0.0029 & 7 \\
1366.01 & 286.358063 & 42.406509 & 15.368 	& 29 $\pm$ 1 & 88.9 $\pm$ 0.1 & 0.031 $\pm$ 0.003 & 19.256 $\pm$  0.019 & 1 \\
1366.02 & 286.358063 & 42.406509 & 15.368 	& 47 $\pm$ 12 & 88.7 $\pm$ 0.5 & 0.05 $\pm$ 0.02 & 54.156 $\pm$  0.021 & - \\
1426.01 & 283.209167 & 48.777641 & 14.232 	& 48.3 $\pm$ 0.6 & 89.67 $\pm$ 0.09 & 0.029 $\pm$ 0.001 & 38.868 $\pm$  0.011 & - \\
1426.02 & 283.209167 & 48.777641 & 14.232 	& 93 $\pm$ 11 & 89.6 $\pm$ 0.2 & 0.065 $\pm$ 0.002 & 74.927 $\pm$  0.011 & 1 \\
1426.03 & 283.209167 & 48.777641 & 14.232 	& 131 $\pm$ 8 & 89.56 $\pm$ 0.05 & 0.12 $\pm$ 0.03 & 150.025 $\pm$  0.013 & - \\
1429.01 & 292.351501 & 48.511082 & 15.531 	& 114 $\pm$ 21 & 89.6 $\pm$ 0.2 & 0.051 $\pm$ 0.003 & 205.914 $\pm$  0.021 & - \\
1474.01 & 295.417877 & 51.184761 & 13.005 	& 50.4 $\pm$ 0.7 & 88.69 $\pm$ 0.05 & 0.25 $\pm$ 0.03 & 69.721 $\pm$  0.032 & 2 \\
1573.01 & 296.846161 & 40.138611 & 14.373 	& 59.4 $\pm$ 0.9 & 89.80 $\pm$ 0.06 & 0.045 $\pm$ 0.001 & 24.8093 $\pm$  0.0058 & 6\\
1574.01 & 297.916870 & 46.965130 & 14.600 	& 60 $\pm$ 10 & 89.3 $\pm$ 0.2 & 0.067 $\pm$ 0.003 & 114.7356 $\pm$  0.0079 & - \\
1574.02 & 297.916870 & 46.965130 & 14.600       & 48 $\pm$ 7 & 88.9 $\pm$ 0.2 & 0.036 $\pm$ 0.001 & 191.29 $\pm$  0.17 & - \\
1873.01 & 295.809296 & 40.008511 & 15.674 	& 63.9 $\pm$ 0.4 & 89.90 $\pm$ 0.02 & 0.045 $\pm$ 0.002 & 71.3106 $\pm$  0.0087 & - \\
2672.01 & 296.132812 & 48.977402 & 11.921 	& 80 $\pm$ 12 & 89.5 $\pm$ 0.2 & 0.051 $\pm$ 0.003 & 88.508 $\pm$  0.013 & - \\
2672.02 & 296.132812 & 48.977402 & 11.921 	& 72.2 $\pm$ 0.3 & 89.92 $\pm$ 0.01 & 0.0303 $\pm$ 0.0009 & 42.9933 $\pm$  0.0042 & - \\
    \hline
    \end{tabular}
\end{table*}

\subsection{Primary transit fitting and detrending of ground-based data}
\label{sec:PTF2}

Once DIP$^2$OL returns the photometric light curve and the associated
detrending quantities, the computation of ground-based TTVs
begins. First, we convert the time-axis, originally given in Julian
dates, to Barycentric Julian dates using \citet{Eastman2010} web
tool\footnote[1]{http://astroutils.astronomy.ohio-state.edu/time/utc2bjd.html}. To
do so, we make use of the celestial coordinates of the star, the
geographic coordinates of the site, and the height above sea level.
Throughout this work, our model comprises a primary transit times a
detrending component. Thus, to compute TTVs we carry out a more
refined detrending of the light curves rather than just a
time-dependent polynomial. For the detrending model we consider a
linear combination of seeing, airmass, (x,y) centroid positions of the
target and of the reference stars, integrated counts over the selected
photometric aperture and the (x,y) centroid positions of the master
flat field and the master dark frames, when available, and integrated
sky counts for the selected sky ring \citep[see e.g.,][for a similar
  approach in the detrending strategy]{Kundurthy2013,Becker2013}. Due
to the nature of the data the exact time at which the mid-transits
will occur are in principle unknown, or known but with a given
certainty. Some photometric observations could actually have been
taken outside the primary transit occurrence. As a consequence, we
have to be extremely careful not to over-fit our data. In order to
choose a sufficiently large number of fitting parameters we take into
consideration the joint minimization of four statistical indicators:
the reduced-$\chi^2$ statistic, $\chi^2_{red}$, the Bayesian
Information Criterion, \mbox{BIC = $\chi^2$ + k ln(Q)}, the standard
deviation of the residual light curves enlarged by the number of
fitting parameters, \mbox{$\sigma_{res}\times$k}, and the Cash
statistic \citep{Cash1979}, \mbox{Cash = 2$\sum_{i=1}^{Q}{M_i -
    D_i*ln(M_i)}$}, being M the model and D the data. For the BIC, k
is the number of fitting parameters, and $\chi^2$ is computed from the
residuals, obtained by subtracting to the synthetic data the best-fit
model. For the BIC and Cash, Q is the number of data points per light
curve. The full detrending model, DM, has the following expression:

\begin{equation}
  \begin{split}
  \mathrm{DM(t) = c_0 + c_1 \cdot \hat{\chi} + c_2\cdot \hat{S} +} \,\\
  \mathrm{\sum_{i = 1}^{N+1} bg_i\cdot BG_i + fc_i \cdot FC_i  + dk_i \cdot DK_i + x_i \cdot X_i + y_i \cdot Y_i}\,
  \end{split}
\end{equation}

\noindent Here, N+1 denotes the total number of target and reference
stars, $\hat{S}$ and $\hat{\chi}$ correspond to seeing and airmass,
respectively. $X_i$ and $Y_i$ are the (x,y) centroid positions. FC$_i$
and DK$_i$ are the integrated flat and dark counts in the chosen
aperture, respectively, and BG$_i$ correspond to the background
counts. The coefficients of the detrending model are $c_0, c_1, c_2$,
and $bg_i, fc_i, dk_i$ and $x_i, y_i$, with \mbox{i = 1, N+1}. Using a
linear combination of these components simplifies the computation of
the detrending coefficients that accompany them by means of simple
inversion techniques. Rather than using the full detrending model to
clean the data from systematics and potentially over-fit the data, we
evaluate sub-models of it (this is, a linear combination of some of
the detrending components). Typical detrending functions would have
the following expression:

\begin{eqnarray}
\mathrm{DM_0 = c_0}\,, \nonumber \\
\mathrm{DM_1 = c_0 + c_1\hat{\chi}}\,, \nonumber  \\
\mathrm{DM_2 = c_0 + c_1\hat{\chi} + c_2 \hat{S}}\,, \nonumber \\
\mathrm{DM_3 = c_0 + c_2 \hat{S}}\,, \nonumber \\
\mathrm{DM_4 = c_0 + \sum_{i=1}^{N+1} bg_i\cdot BG_i}\,, \nonumber \\ 
\mathrm{DM_5 = c_0 + c_1\hat{\chi} + \sum_{i=1}^{N+1} bg_i\cdot BG_i}\,, \nonumber \\
\dotsm \nonumber \\
\mathrm{DM_{14} = c_0 + c_1\hat{\chi} + c_2 \hat{S} + \sum_{i=1}^{N+1} bg_i\cdot BG_i + x_i\cdot X_i + y_i\cdot Y_i}\,. \nonumber \\
\dotsm \nonumber \\
\label{eq:DM}
\end{eqnarray}

\noindent DIP$^2$OL considers a total of 56 sub-models, depending on
the availability of calibrations. Usually, the noise in the data is
correlated with airmass, (x,y) centroid positions and integrated flat
counts, while the dependency with seeing strongly depends on the
photometric quality and stability of the particular night. Therefore,
these 56 sub-models are constructed solely from how we think the
systematics impact the data. Although all possible combinations should
be tested, this is computationally expensive, specially considering
that a differential light curve can be constructed averaging 20-30
reference stars (i.e., N = 20-30).

\begin{figure}[ht!]
  \centering
  \includegraphics[width=.5\textwidth]{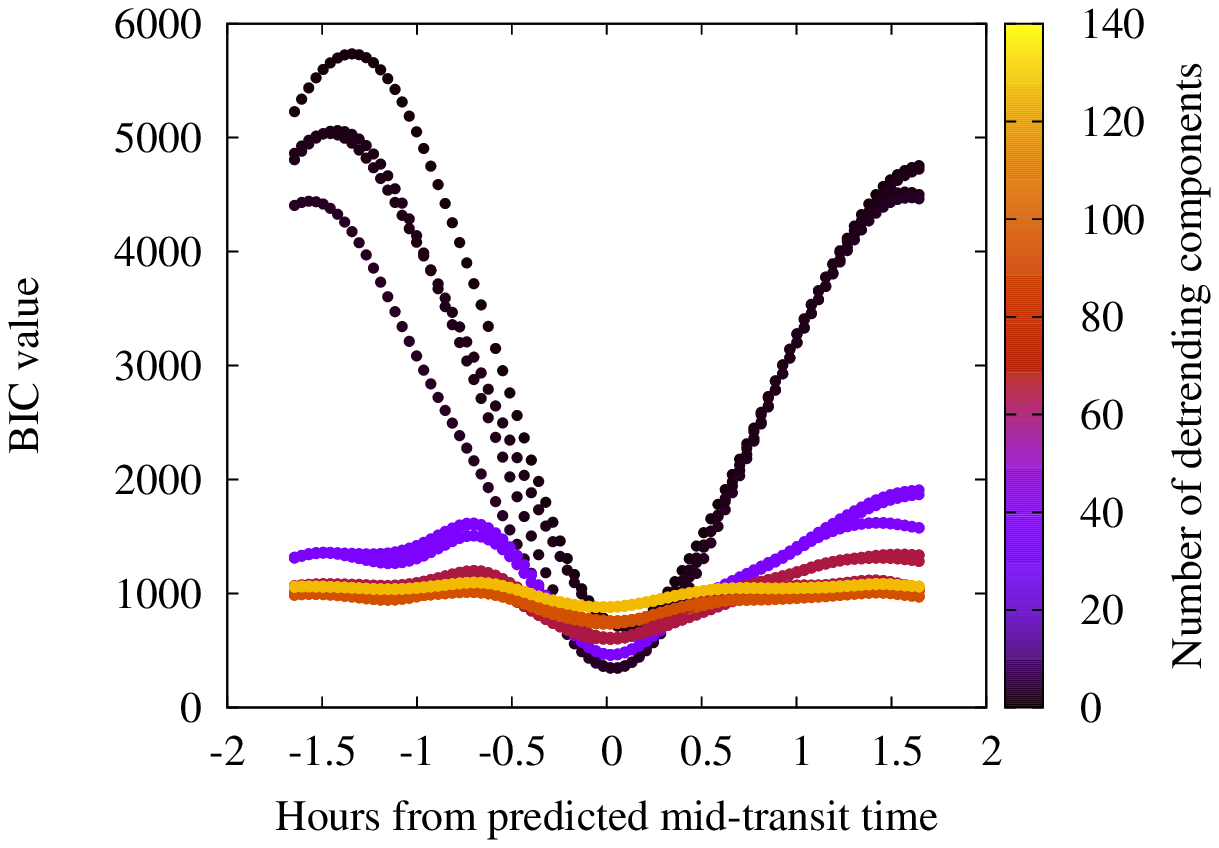}
  \includegraphics[width=.5\textwidth]{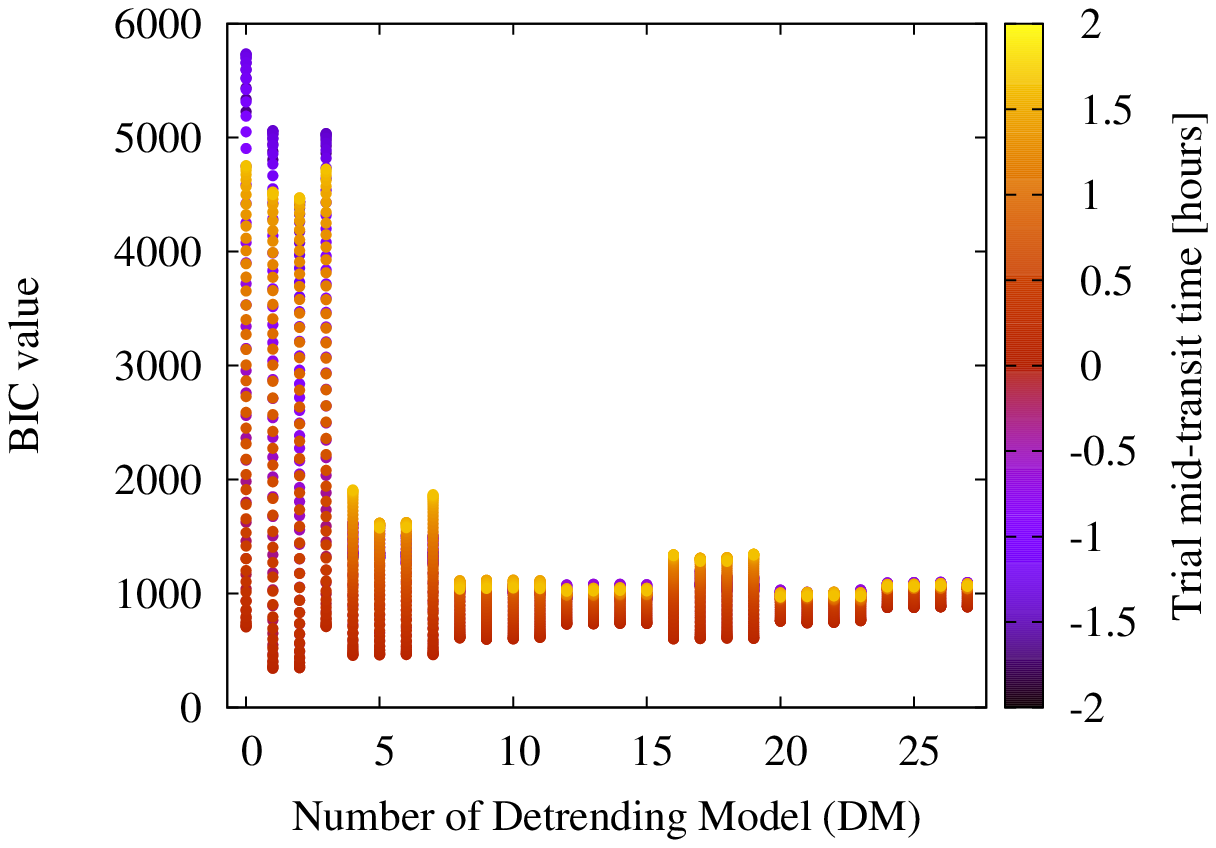}
  \caption{\label{fig:BIC_behav} {\it Top:} BIC values as a function
    of trial mid-transit times. Color-coded are the number of
    detrending components for each one of the detrending models
    (sub-models). {\it Bottom:} BIC values as a function of the
    detrending model, DM. Numbers are in agreement with the labels on
    Eq.~\ref{eq:DM}. Color-coded are the trial $\mathrm{T_0}$'s.}
\end{figure}

To determine the detrending sub-model best matching the residual noise
in the data, we first create an array of trial $\mathrm{T_0}$'s around
the predicted mid-transit time, covering the $\pm$T$_{\mathrm{dur}}$
space and respecting the cadence of the observations. This takes care
of the uncertainty in the knowledge of the mid-transit times, since
typical errors in the predictions of transits with large TTVs can
increase up to 40-50 minutes, in some cases even more. For each one of
these trial $\mathrm{T_0}$'s and each one of the sub-models we compute
the previously mentioned four statistics. In principle, if a given
trial $\mathrm{T_0}$ is close to the true mid-transit time, then
around this $\mathrm{T_0}$ all the sub-models should minimize the four
statistics. To illustrate this, Figure~\ref{fig:BIC_behav}, {\it top},
shows how the BIC changes as a function of the trial $\mathrm{T_0}$,
for all the possible sub-models (28 in this case, since dark frames
were unavailable). For this example, we analyzed the transit
photometry of \mbox{KOI-0760.01} taken with the 2.5~m Nordic Optical
Telescope. Color-coded are the number of detrending
components. Figure~\ref{fig:BIC_behav}, {\it bottom}, shows the
dependency of the BIC with the sub-models (i.e., detrending models,
DM). The numbers on the abscissa are in concordance with the indices
in Eq.~\ref{eq:DM}. Color-coded are the trial $\mathrm{T_0}$'s. For
this data set, the BIC minimizes at DM$_2$. As a consequence, the data
do not correlate with the integrated flat counts, nor the centroid
position. This is actually what we expect, since the Nordic Optical
Telescope has an outstanding guiding system that can keep stars within
the same pixels for hours.

\begin{figure}[ht!]
  \centering
  \includegraphics[width=.5\textwidth]{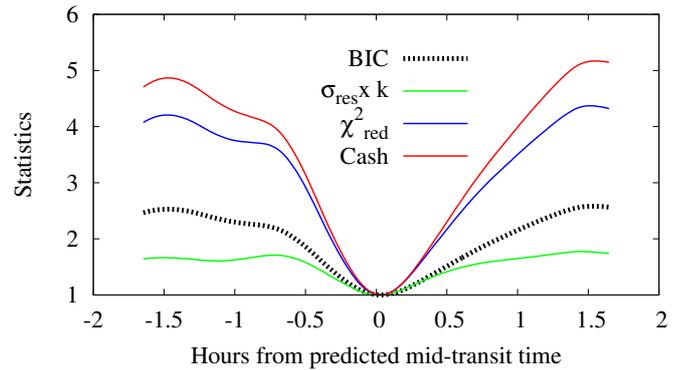}
  \caption{\label{fig:statistics}The four statistics used to assess
    the number of detrending components and the starting mid-transit
    time, obtained analyzing \mbox{KOI-0760} data. Their values have
    been normalized and scaled to allow for visual comparison.  The
    thick dashed black line corresponds to the time-averaged BIC
    statistics, as shown in the top panel of
    Figure~\ref{fig:BIC_behav}.}
\end{figure}

Then, we make use of the minimization of the time-averaged statistics
(that is, the statistics averaged within each one of the
$\mathrm{T_0}$'s) to determine the starting value of the mid-transit
time that will be used in our posterior transit fitting (see
Figure~\ref{fig:statistics}). This is a more robust approach than
simply computing the absolute minimum value of the statistics, since
these could be produced by chance. Finally, with this mid-transit time
fixed we re-compute the transit model and re-iterate over all the
detrending models to choose the one that minimizes the averaged
statistics.

For the transit fitting instance we use a quadratic limb darkening law
with quadratic limb darkening values computed as described in
\cite{vonEssen2013}, for the filter band matching the one used during
the observations and for the stellar fundamental parameters closely
matching the ones of the KOIs. Rather than considering the orbital
period, the inclination, the semi-major axis and the planet-to-star
radius ratio as fixed parameters to the values given by the NASA
Exoplanet Archive or the values derived in
Table~\ref{tab:KOIs_params}, we use a Gaussian probability
distribution which mean and standard deviation equals the values
obtained in Section~\ref{sec:PTF1}, and we fit all of them
simultaneously to the detrending model and the mid-transit time. The
inclination, semi-major axis and planet-to-star radius ratio are
fitted only if the light curves show complete transit coverage. If
not, we consider them as fixed to the values reported in
Table~\ref{tab:KOIs_params}, and we fit only the mid-transit time. At
each MCMC step the transit parameters change. Therefore, for each
iteration we compute the detrending coefficients with the previously
mentioned inversion technique. To fit KOINet's ground-based data we
produce 5$\times$10$^6$ repetitions of the MCMC chains, we discard the
first 20\%, and we compute the mean and standard deviation
\mbox{(1-$\sigma$)} of the posterior distributions of the parameters
as best-fit values and uncertainties, respectively. To check for
the convergence of the chains, we divide the remaining 80\% in four, and
we compute mean and standard deviations of the priors within each
20\%. We consider that the chains converged if all the values are
consistent within 1-$\sigma$ errors. Finally, we visually inspect the
posterior distributions and their correlations.

To provide reliable error bars on the timing measurements we evaluate
to what extent our photometric data are affected by correlated
noise. To this end, following \cite{Carter2009} we compute residual
light curves by dividing our photometric data by the best-fit transit
and detrending models. From the residuals, we compute the $\beta$
factors as specified in \cite{vonEssen2013}. Here, we divide each
residual light curve into M bins of N averaged data points. If the
data are free of correlated noise, then the noise within the residual
light curves should follow the expectation of independent random
numbers:

\begin{equation}
  \hat{\sigma}_N = \sigma_1 N^{-1/2}[M/(M-1)]^{1/2}\ .
\end{equation}

\noindent Here, $\sigma_1$ corresponds to the scatter of the unbinned
residuals, and $\sigma_N$ to the variance of the binned data:

\begin{equation}
  \sigma_N = \sqrt{\frac{1}{M}\sum_{i = 1}^{M}(<\hat{\mu}_i> - \hat{\mu}_i)^2}\ ,
\end{equation}

\noindent In the equation, the mean value of the residuals per bin is
given by $\hat{\mu_i}$, and $<\hat{\mu_i}>$ is the mean value of the
means. In the presence of correlated noise, $\sigma_N$ should differ
by the factor $\beta_N$ from $\hat{\sigma}_N$. Therefore, we compute
$\beta$ by averaging the $\beta_N$'s obtained over time bins close to
the duration of ingress, which is in turn computed from the previously
determined best-fit transit parameters. In particular, we consider
time bins as large as 0.8, 0.9, 1, 1.1, and 1.2 times the duration of
ingress. If $\beta$ is larger than 1, we enlarge the photometric error
bars by this value, and we carry out the MCMC fitting in the exact
same fashion as previously explained. We conclude by visually
inspecting the data, the detrending model, and the best-fit transit
model.

\section{KOINet's achieved milestones}
\label{sec:MS}

\subsection{Photometric data presented in this work}

Here we present seven new primary transit observations of
\mbox{KOI-0760.01}, \mbox{KOI-0410.01}, \mbox{KOI-0525.01} and
\mbox{KOI-0902.01}. These were taken between May, 2014 and July,
2015. Table~\ref{tab:obs} shows the basic photometric characteristics
of the data, such as the photometric precision, the cadence and number
of data points per light curve, the transit coverage and the dates at
which the KOIs were observed. The last column of Table~\ref{tab:obs}
shows their corresponding mid-transit times, along with 1-$\sigma$
uncertainties. The KOIs presented in this work were chosen to
exemplify the need for a network such as KOINet, as described in the
following sections.

We would like to emphasize that the purpose of this paper is to show
the potential of KOINet. Therefore, we present the TTV observations,
together with preliminary models. For individual cases (e.g.,
\mbox{KOI-0410.01}) a more in-depth analysis is also given.

\begin{table*}[ht!]
  \caption{\label{tab:obs} Most relevant parameters obtained from our
    observations. From left to right: the date corresponding to the
    beginning of the local night in years, months and days
    (yyyy.mm.dd), the name of the observed KOI, the telescope
    performing the observations, the standard deviation of the
    residual light curves in parts-per-thousand (ppt), $\sigma_{res}$,
    the number of data points per light curve, N, the average cadence
    in seconds, CAD, the total observing time, T$_{tot}$, in hours,
    and the airmass range, $\chi_{min,max}$, showing minimum and
    maximum values, respectively, the transit
    coverage, TC (a description of the transit coding is detailed in
    the footnote of this table) and the derived mid-transit times, in BJD$_{TDB}$.}  \centering
  \begin{tabular}{l c c c c c c c c c}
    \hline \hline
    Date       & Telescope   &   Name           &  $\sigma_{res}$  &    N    &   CAD   &   T$_{tot}$   &   $\chi_{min,max}$   &     TC    &    TTVs $\pm$ 1-$\sigma$  \\
    yyyy.mm.dd &             &                  &  (ppt)          &         &  (sec)  &   (hours)    &             &          &    BJD$_{TDB}$    \\
    \hline
    2014.06.24 &  IAC 0.8m    & KOI-0902.01      & 1.8             & 114     &  210    &   6.67       & 1.03,1.63  & --BEO  & 2456832.43925  $\pm$ 0.0063 \\
    2014.08.28 &  ARC 3.5m    & KOI-0525.01      & 0.7             & 135     &   60    &   2.26       & 1.02,1.06  & OIBE- & 2456897.71506  $\pm$ 0.0045\\
    2014.10.03 &  NOT 2.5m    & KOI-0760.01      & 1.4             & 302     &   59    &   5.00       & 1.07,2.50  & OIBEO & 2456934.43353  $\pm$ 0.0010\\
    2014.10.12 &  NOT 2.5m    & KOI-0410.01      & 1.0             & 178     &   81    &   4.01       & 1.06,2.29  & OIBEO & 2456943.41142  $\pm$ 0.0018\\
    2014.11.03 &  YO 2.4m     & KOI-0410.01      & 1.4             & 128     &  110    &   3.91       & 1.10,2.78  & OIBEO & 2456965.06535  $\pm$ 0.0020 \\            
    2015.07.06 &  CAHA 2.2m   & KOI-0410.01      & 3.7             & 670     &  41     &   7.75       & 1.00,1.50  & OIBEO & 2457210.43420  $\pm$ 0.0031\\
    2015.07.06 &  NOT 2.5m    & KOI-0410.01      & 0.8             & 142     &  95     &   3.77       & 1.02,1.51  & -IBEO & 2457210.43652  $\pm$ 0.0022\\
    \hline
  \end{tabular}
  \tablefoot{Acronyms for the telescopes are as specified in
    Section~\ref{subsec:IC}. The letter code specifying the transit
    coverage during each observation is the following: O: out of
    transit, before ingress. I: ingress. B: flat bottom. E: egress. O:
    out of transit, after egress.}
\end{table*}

Most of the KOIs that are members of KOINet's follow-up are in wide
orbits. In particular, the average orbital period of these KOIs is
around 65 days, while the largest orbit corresponds to 335 days. As a
consequence, most of the transits last several hours, with an average
value of about \mbox{6 $\pm$ 3} hours. Thus, it is challenging to
cover the full transit from the ground. Normally, incomplete transit
coverage has a large impact in the determination of the mid-transit
times \citep[see e.g.][]{Winn2008,Gibson2009}. However, in our case
this is alleviated by the prior knowledge of the orbital parameters
given by Kepler photometry. While long cadence Kepler data were
obtained averaging images each 30 minutes, the collected ground-based
observations have a cadence between some seconds to a few
minutes. Although our Earth's atmosphere and typical ground-based
instrumental imperfections considerably decrease the photometric
precision when compared to Kepler data, we gain in sampling rate and,
thus, in timing precision, given the prior knowledge of the transit
parameters.

\subsection{Timing precision: KOI-0760.01}

Figure~\ref{fig:kois_data}(a), shows a primary transit of
\mbox{KOI-0760.01} and its corresponding TTVs. The target is a
Neptune-sized planet candidate in a $\sim$5 day orbit, showing TTVs of
small amplitude ($\sim$140 seconds). The TTV period is of around 3.5
years and, as a consequence, has been roughly covered by the Kepler
data. The TTVs show also a scatter of $\sim$2 minutes. To include a
KOI into KOINet's follow-up list, it has to fulfill specific
conditions. One of them is to present TTVs larger than (or about) two
minutes. \mbox{KOI-0760.01} is close to the lower end of this
limit. Compared with Kepler averaged errors in the mid-transit times
($\sim$80 seconds) ground-based data collected by KOINet delivered a
timing precision of the same quality. The left panel of
Figure~\ref{fig:kois_data}(a), shows a detrended transit light curve
in green, the best fit model in black continuous line, and the raw
data and detrended model in blue points and black continuous line,
respectively. For a better visualization, these were arbitrarily
shifted. The figure on the right shows Kepler and ground-based TTVs in
red diamonds. Red and green-shaded areas indicate Kepler coverage and
the ground-based, 2014 observing season, respectively.

\subsection{Photometric precision: KOI-0525.01}
\label{sec:PA}

Another strong limit set by the nature of ground-based observations is
given by the transit depth. For Kepler targets with TTVs, this is
aggravated by the faintness of the stars, the length of the transits,
and the uncertainty of the transit occurrence. Before starting the
observing seasons we set an ambitious lower limit in the transit depth
of $\sim$1 ppt, with the final cut given by the TTV amplitude and the
transit duration. To maximize the transit detection, we assigned these
transits to the largest telescopes. One successful example of our
observing strategy is given by \mbox{KOI-0525.01}. The KOI is labeled
as an exoplanet candidate, and beside the small ($\sim$1 ppt) transit
depth, it has an orbital period of $\sim$11.5 days, and a transit
duration of $\sim$2.25 hours, facilitating observations of complete
transits from the ground. The TTV periodicity shown by Kepler data is
longer than four years, so our observations will help to set
constraints on the nature of the KOI. Figure~\ref{fig:kois_data}(b),
shows the ground-based light curve obtained with the \mbox{ARC 3.5m}
telescope. We detected the transit and a turn-over in the TTVs of
KOI-0525.01, shown in the right side of Figure~\ref{fig:kois_data}(b).
With the addition of the holographic diffuser on the new ARCTIC imager
on the \mbox{ARC 3.5 m} Telescope at Apache Point Observatory, we may
be able to achieve even higher photometric precision in future
observations \citep{Stefansson2017}.

\subsection{Relevance for ground-based follow-ups: KOI-0902.01}

When the TTV periodicity is not fully sampled, it can occur that the
quickly increasing separation between the two most likely scenarios
(sinusoidal and parabolic predictions) prevents us from finding future
transits of these KOIs. For some KOIs, like \mbox{KOI-0410.01} (see
next Section) the difference between these two extreme scenarios is of
the order of a couple minutes and will be well contained below some
hours in the upcoming years. However, for some other KOIs the current
difference is larger than the duration of an observing
night. Therefore, if follow-up campaigns are not carried out in time
it will be extremely expensive to detect the transits of these planet
candidates again. A clear example of this is given by
\mbox{KOI-0902.01}. The exoplanet candidate is in an $\sim$84 days
orbit, and the transit duration is of $\sim$6.7 hours, making the full
observation of a single transit quite challenging from only one
observing site. Figure~\ref{fig:kois_data}(c), shows one of the
transits observed by the \mbox{IAC 0.8m} telescope. The transit
($\Delta F\sim$1\%) has been clearly detected by KOINet. On the right panel,
the TTVs of Kepler data plus the ground-based detected mid-time can be
seen.

\subsection{Turn-over of TTVs: KOI-0410.01}
\label{sec:TTVTO}

\begin{figure*}
  \centering
  (a)

  \includegraphics[width=.45\textwidth]{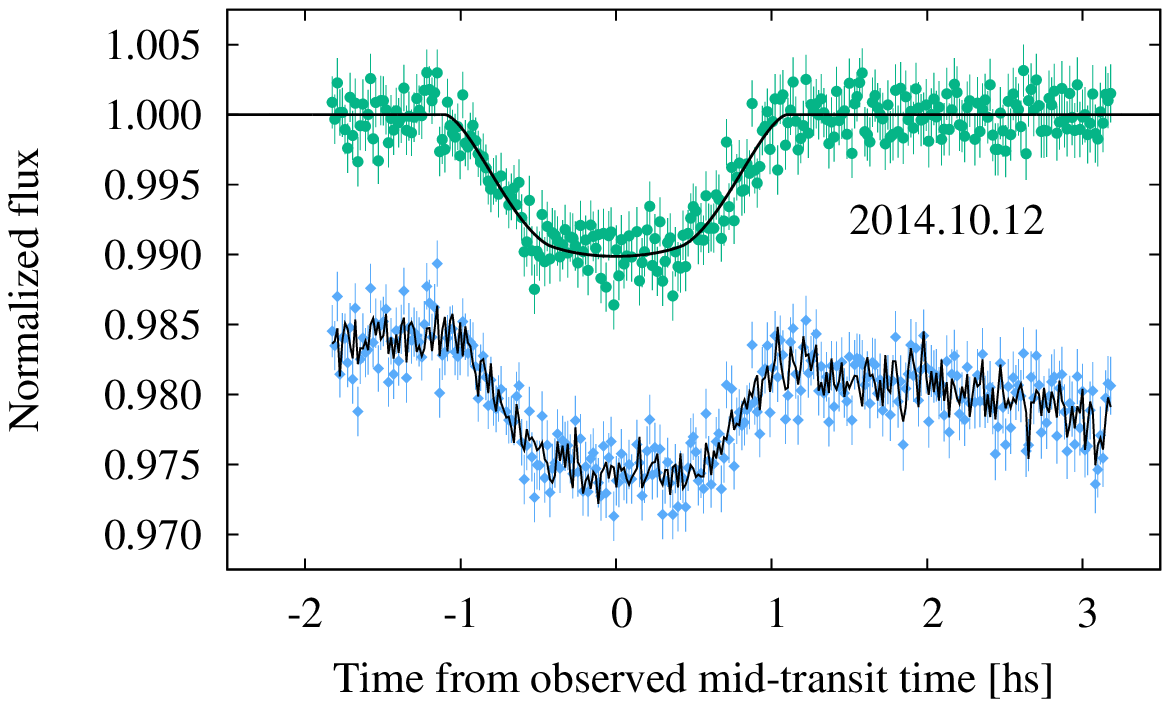}%
  \includegraphics[width=.45\textwidth]{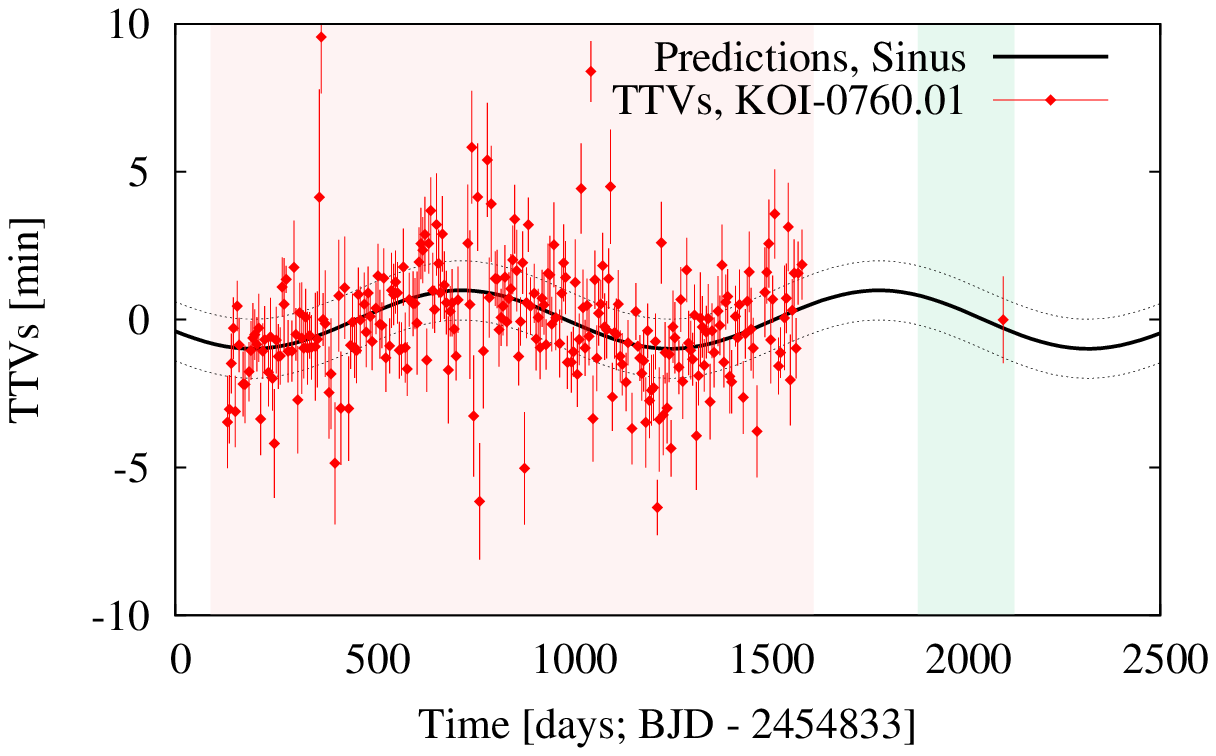}

  (b)

  \includegraphics[width=.45\textwidth]{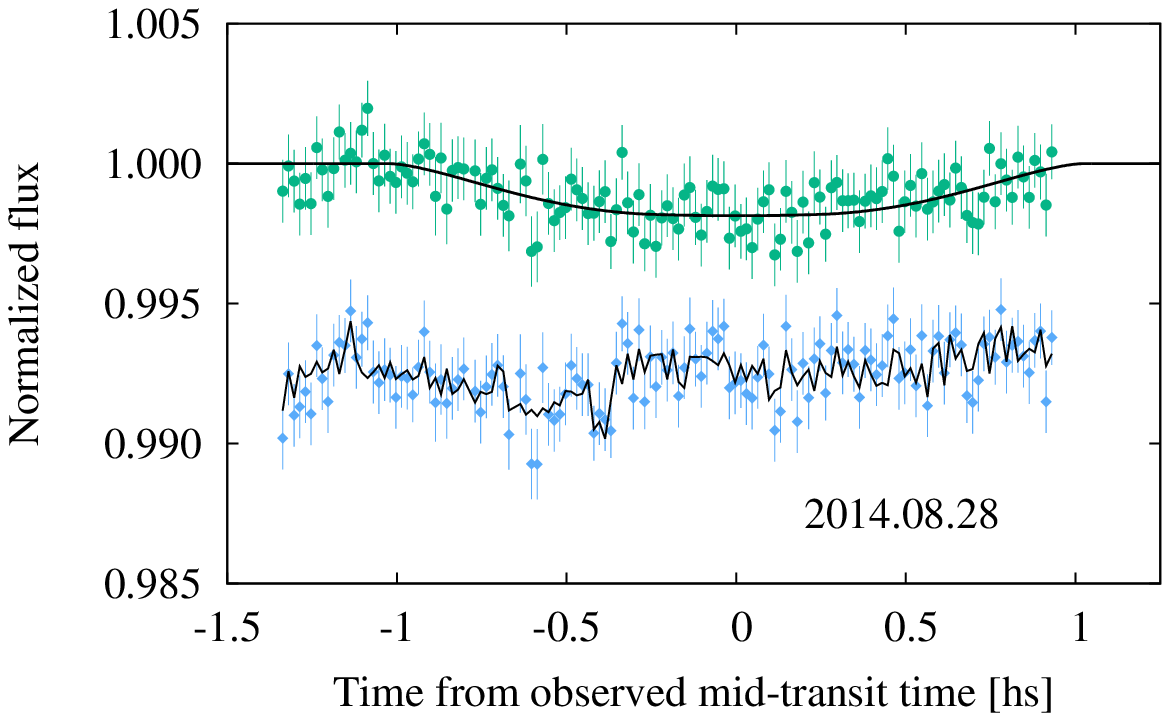}%
  \includegraphics[width=.45\textwidth]{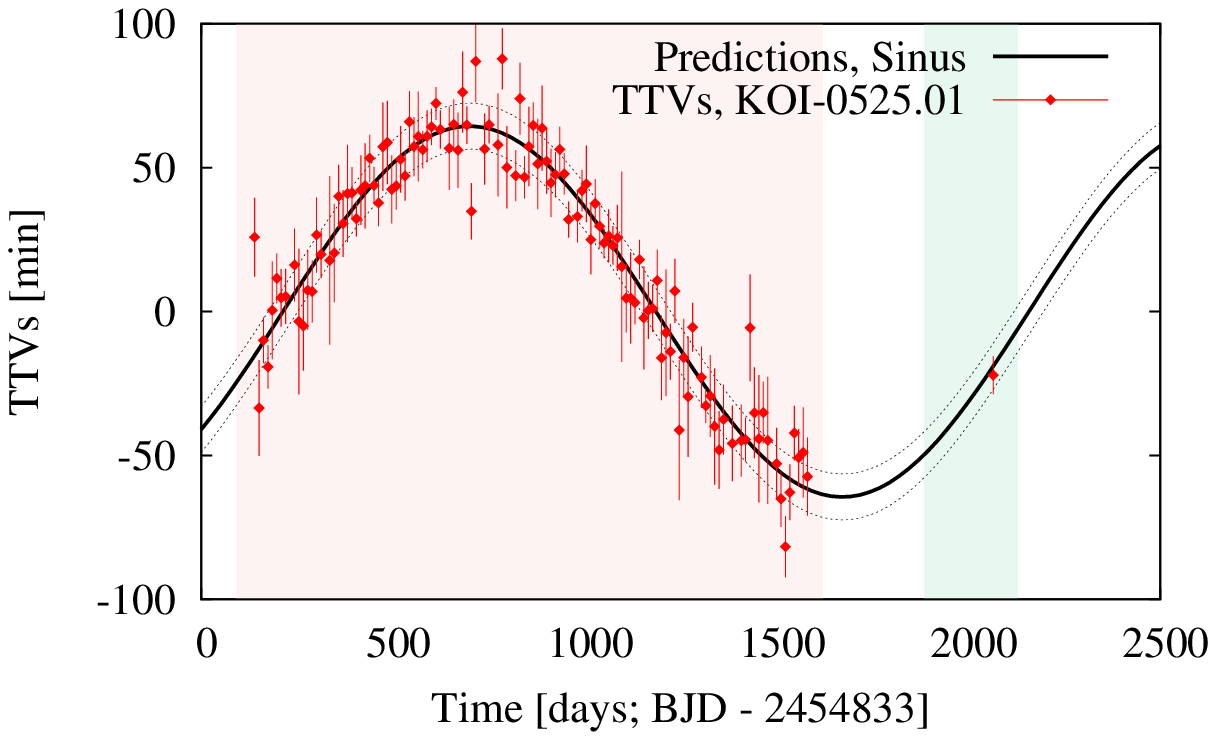}

  (c)

  \includegraphics[width=.45\textwidth]{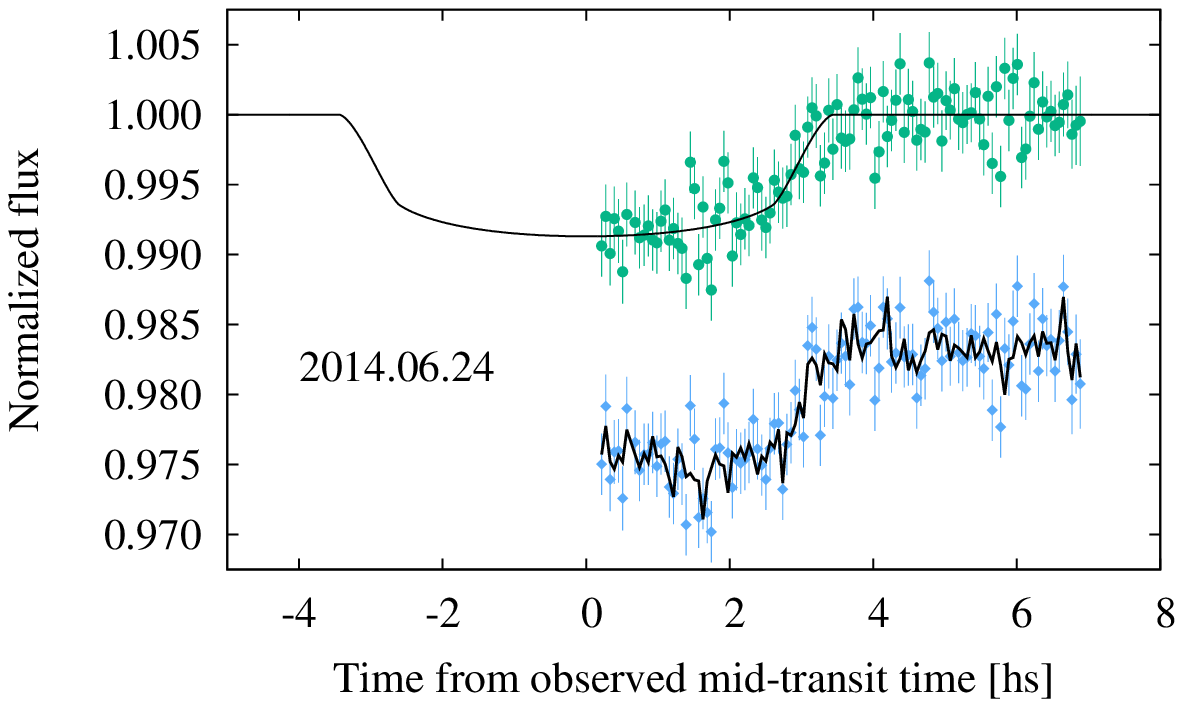}%
  \includegraphics[width=.45\textwidth]{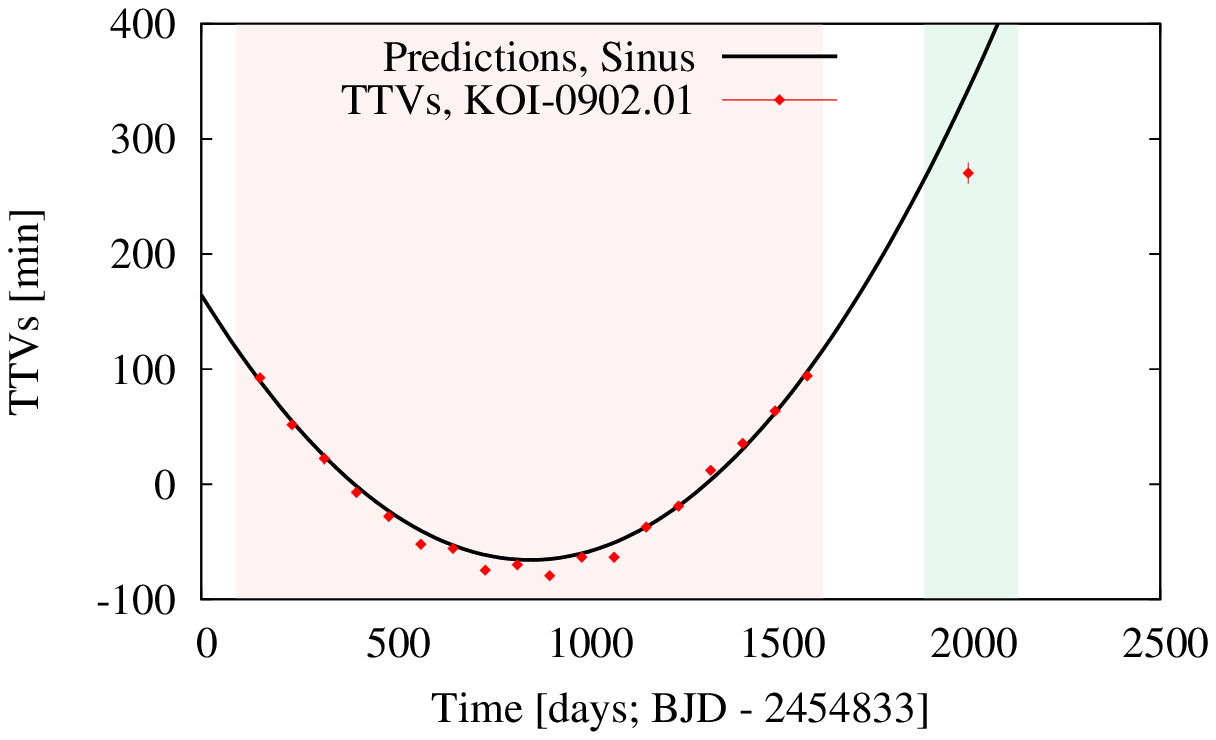}

  (d)

  \includegraphics[width=.45\textwidth]{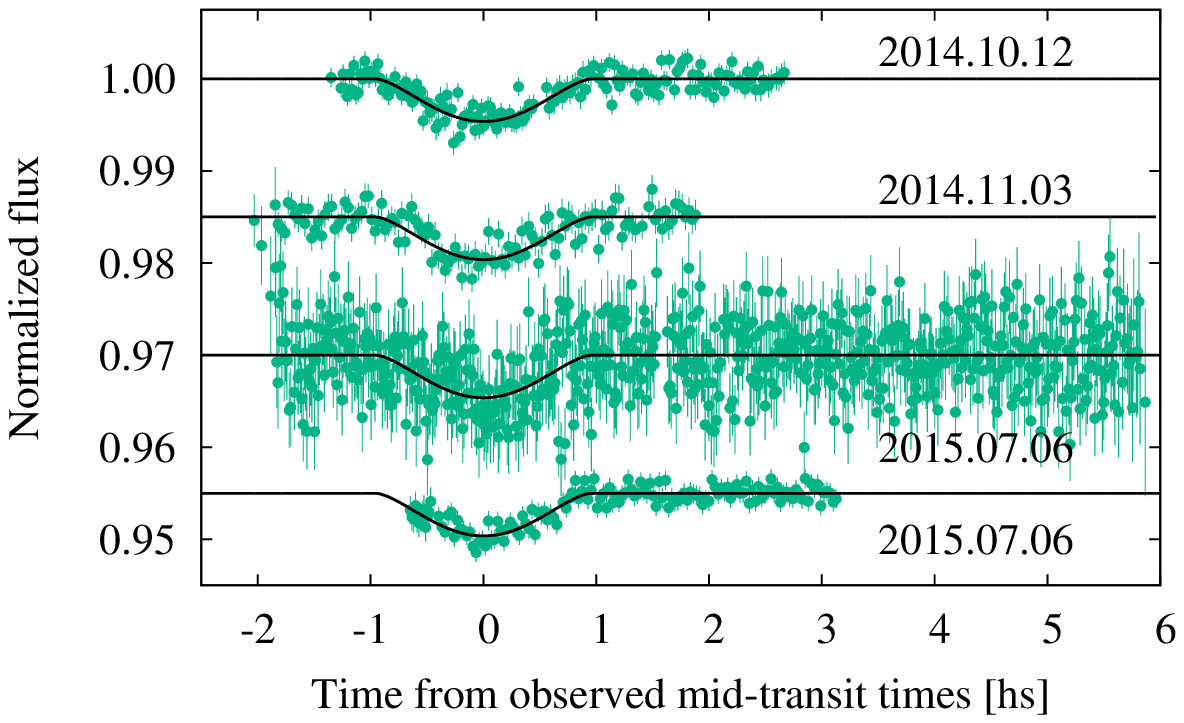}%
  \includegraphics[width=.45\textwidth]{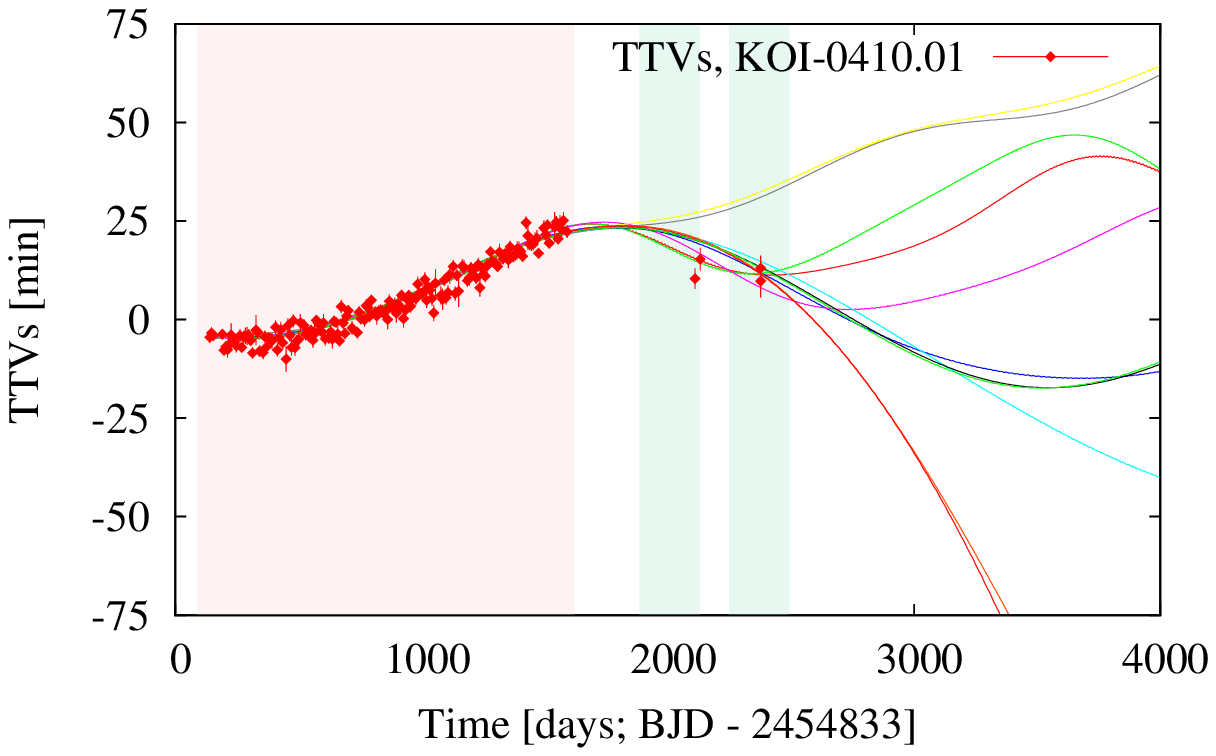}

 \caption{\label{fig:kois_data} {\it Left figures:} From top to
   bottom, transit observations of \mbox{KOI-0760.01},
   \mbox{KOI-0525.01}, \mbox{KOI-0902.01}, and \mbox{KOI-0410.01}. The
   data are plotted in hours with respect to their best-fit
   mid-transit times. In all cases green circles show detrended data,
   along with the best-fit model as black continuous lines. For the
   first three cases only, the raw data are shown as blue diamonds,
   plus the detrending times transit models are also overplotted as
   black lines, shifted down for a better visual comparison, to
   exemplify our detrending strategy. {\it Right:} Kepler and
   ground-based TTVs in red diamonds, for the KOIs displayed on the
   left. Continuous and dashed black and color lines indicate the
   predictions and uncertainties, respectively. While the red shaded
   area represents Kepler observing time, the green shaded areas
   represent the 2014 and 2015 observing seasons (March til October).}
\end{figure*}

In the NASA Exoplanet Archive, more specifically within the Data
Validation Reports (DV), \mbox{KOI-0410.01} has been identified as a
false positive. Most of the evidence in favor of this status is based
upon the V-shape of its transits. The flux drops take place each
$\sim$7.2 days, and the transit depth is about 5 ppt. During 2014
and 2015 we have observed four transits of \mbox{KOI-0410.01}. The
characteristics of the data are presented in Table~\ref{tab:obs}. The
combination of Kepler and ground-based TTVs, shown in
Figure~\ref{fig:kois_data}(d), reveal a turn-over in their mid-times.
This, together with the assigned status, motivates a more detailed
study to shed some light into the nature of the system.

To begin with, in the DV report a source of 18.4 Kepler magnitudes has
been detected, approximately nine arcseconds away from the nominal
position of \mbox{KOI-0410}. Therefore, it could be possible that a
background eclipsing binary could be causing the observed flux drops,
there. For well spatially resolved ground-based observations, 9
arcseconds corresponds to several dozens of pixels. To identify the
source, we combined all the observations of our best (sharpest)
night. Once the mentioned source was located, we carried out the usual
differential photometry, choosing different apertures centered on
\mbox{KOI-0410}, both including and excluding the source. As expected,
the increase or decrease in aperture (from 0.9 to 10 arcseconds)
changed the overall photometric accuracy of the differential light
curves. However, it did not change the shape or depth of the
transit. Therefore, as also observed by \cite{Bouchy2011} it is
unlikely that the flux drops are caused by a background binary 9
arcseconds away from \mbox{KOI-0410}.

\begin{figure}[ht!]
  \centering
  \includegraphics[width=.5\textwidth]{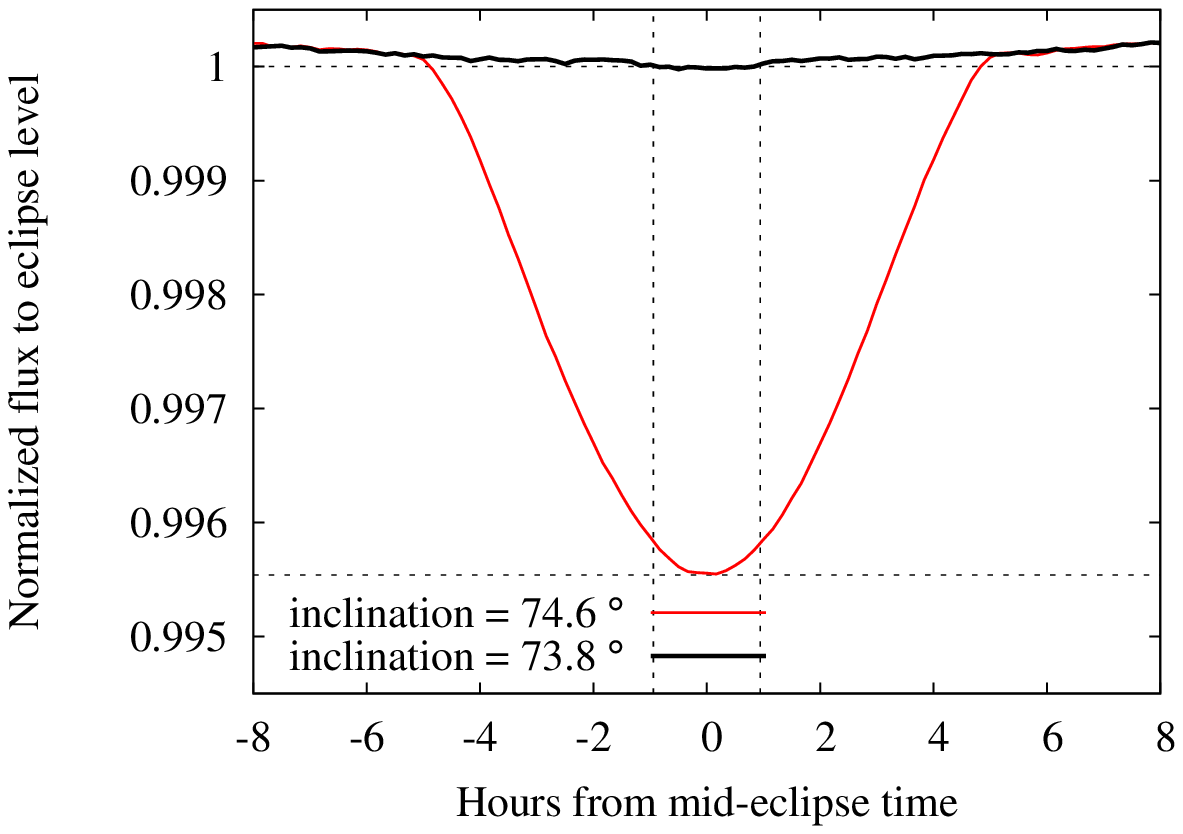}
  \includegraphics[width=.5\textwidth]{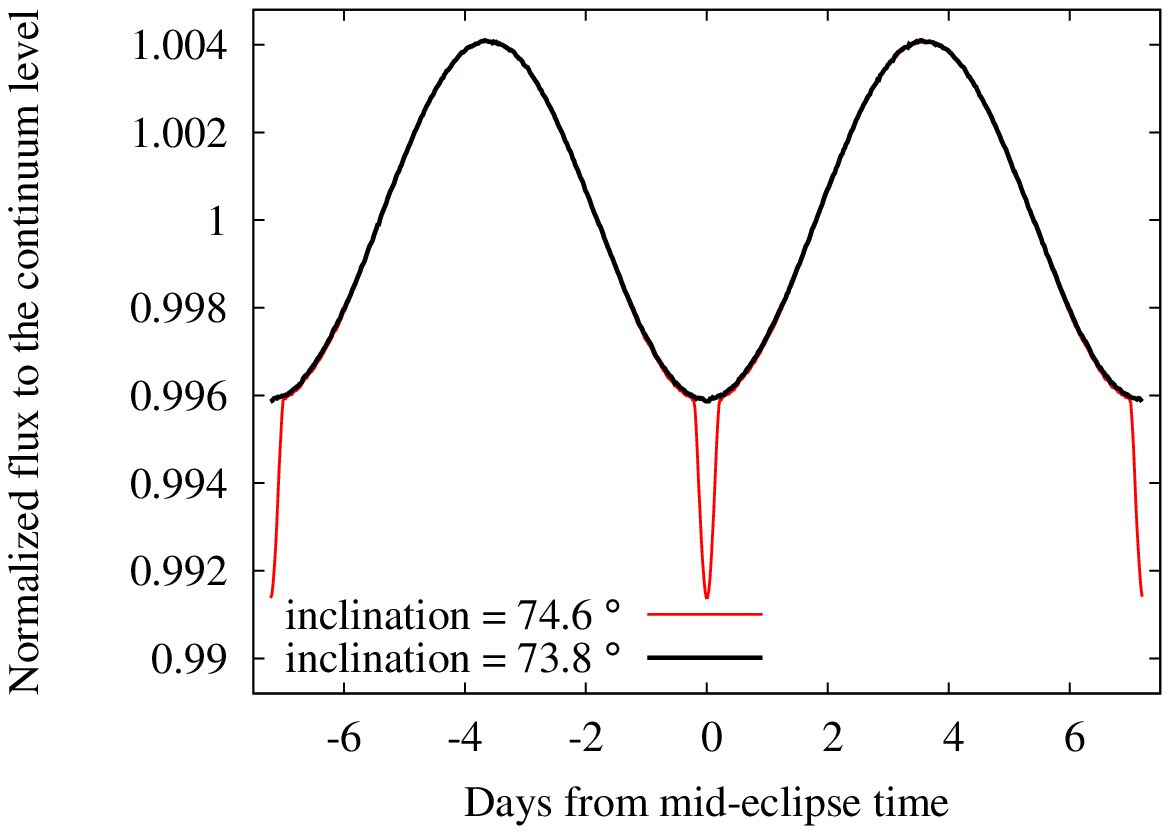}
  \caption{\label{fig:PHOEBE} PHOEBE models for two identical G-type
    stars in a $\sim$14.4-day orbit, for an inclination of
    74.6$^{\circ}$ in red, and 73.8$^{\circ}$ in black continuous
    lines. {\it Top:} zoom-in around the eclipses. Time is given in
    hours. Horizontal and vertical dashed lines indicate the transit
    duration and depth, respectively, for visual inspection. {\it
      Bottom:} complete orbits, highlighting the continuum flux
    variability.}
\end{figure}

Furthermore, we investigated the case in which the flux drops were
caused by a grazing eclipsing binary in a $\sim$7.2 days orbit, or two
similar stars in a $\sim$14.4-day orbit, eclipsing each other each
$\sim$7.2 days. In both cases, the radial velocity shifts caused by
their mutual orbital motion would create a detectable variability, not
observed by \cite{Bouchy2011}. These authors observed \mbox{KOI-0410}
at two opportunities, at orbital phases close to 0.5 and 0.75,
detecting RV shifts inconsistent with the ones expected to be caused
by two orbiting stars. To examine the scenario of two identical stars
in more detail, we simulated two identical G-type stars using the
PHysics Of Eclipsing BinariEs\footnote{http://phoebe-project.org/}
\citep[PHOEBE][]{Prsa2016} code. We assumed stellar radii consistent
with G-type stars, with limb darkening coefficients chosen accordingly
and computed as described in \cite{vonEssen2013}, and we changed the
inclination to match both the eclipse duration and depth. On one hand,
for an inclination of 74.6 degrees we matched the eclipse depth
($\sim$5 ppt, 4460.6 $\pm$ 17.8 parts per million, as reported in the
NASA Exoplanet Archive). However, the eclipse duration exceeds the
observed one (T$_{\mathrm{dur}}$ = 1.899 hours, NASA Exoplanet
Archive). On the other hand, decreasing the inclination to 73.8
degrees decreases the eclipse duration considerably (less than 2
hours). Nonetheless, the eclipse depth was smaller than $\sim$0.05 ppt
(see Figure~\ref{fig:PHOEBE}, {\it top}). Thus, we could not match
both transit depth and duration of \mbox{KOI-0410.01} to the light
curves produced from PHOEBE models. In addition to this, both models
predict a variability outside eclipse of about $\sim$1\%, which would
be clearly visible in Kepler data (Figure~\ref{fig:PHOEBE}, {\it
  bottom}). After visually inspecting the raw data of the 17 quarters
and finding no variability modulated with the mentioned periodicity
and amplitude, we believe is unlikely that the flux drop is caused by
two identical stars in grazing orbits. In addition, the stars should
be in exact circular orbits to not show odd/even timing differences,
these in turn not observed in the DV report.

\begin{figure}[ht!]
  \centering
  \includegraphics[width=.5\textwidth]{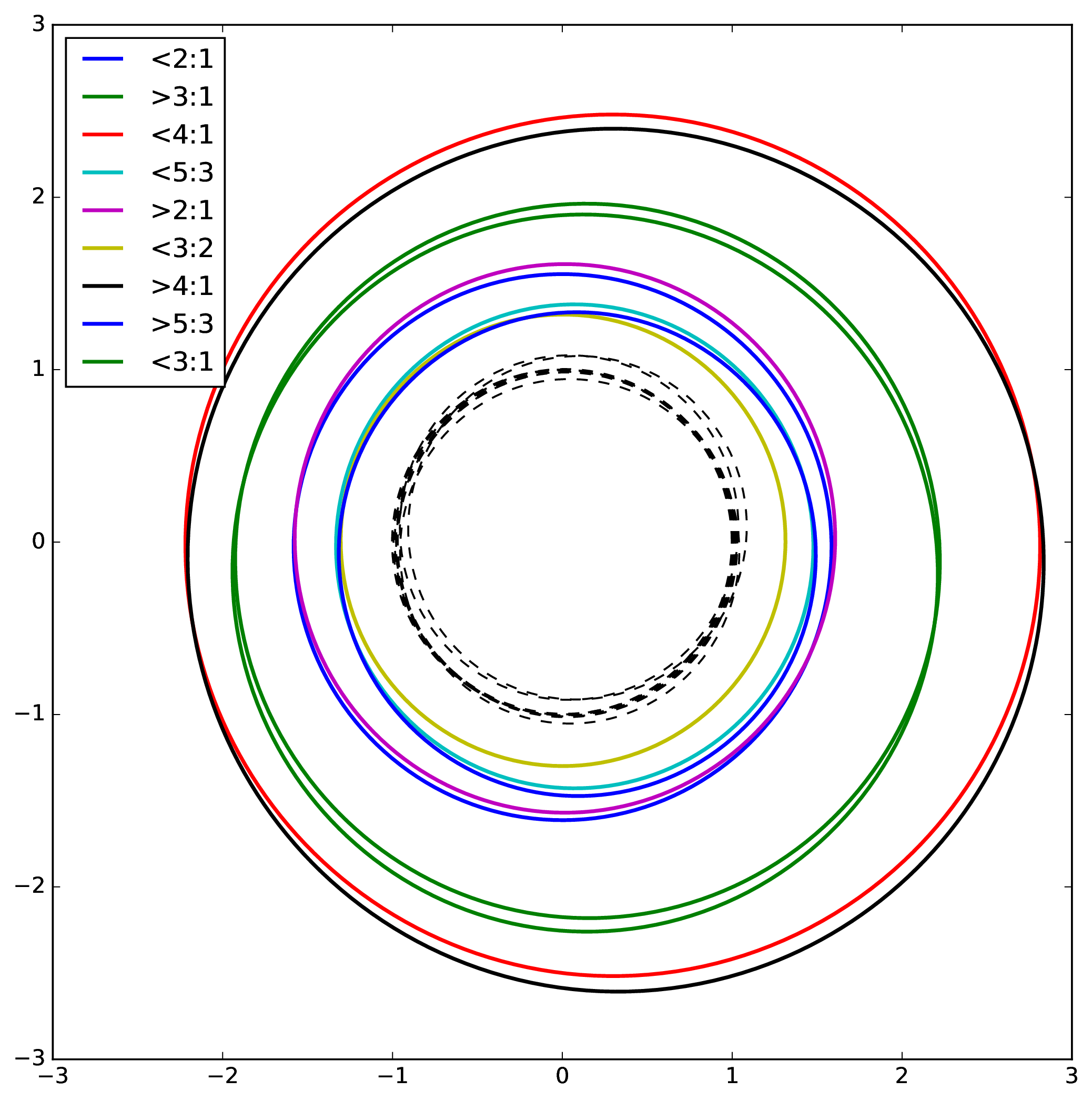}
  \caption{\label{fig:TTVFast} The shapes of the orbits for each one
    of the exterior perturber TTV solutions. The dashed lines
    represent the orbits for the transiting planet in each case. The
    axis units are in terms of the semi-major axis of the transiting
    planet. The exact values for the mass and period ratios are
    specified in Table~\ref{tab:TTVFast}. The solution just interior
    to 3:2 isn't plotted due to large $\chi^2$ values.}
\end{figure}

Finally, we investigated another two scenarios: a
gravitationally-bound hierarchical triple star system, and a
planet-star system. Here we show some initial results assuming a
planet-star system. In this case, the simultaneous MCMC transit
fitting of ground and space-based data allows for a planet
solution. Assuming this, the right panel of
Figure~\ref{fig:kois_data}(d) shows some initial TTV results obtained
from the data acquired during the first and second observing seasons,
together with Kepler data. As the figure shows, additional data
already scheduled for the 2018 observing season will constrain the
solutions even further. The TTV fits were carried out with a novel
version of \texttt{TTVFast} \citep{Deck2014} which utilizes an
symplectic integrator developed by \citet{Hernandez2015} and the
universal Kepler solver of \citet{Wisdom2015}. We computed the TTVs
caused by a single, outer perturbing planet near 5:3, 3:2, 2:1, 3:1,
and 4:1 period ratios (see Figure~\ref{fig:TTVFast} and
Table~\ref{tab:TTVFast}), starting the solution just interior and just
exterior to each resonance with a super-period \citep{Lithwick2012}
corresponding to $\approx 3000$ days, following the analysis by
\citet{Ballard2011} of Kepler-19. We held the mass-ratio of the
transiting planet \mbox{(KOI-0410.01)} to the star fixed at $10^{-5}$,
while we allowed the initial ephemerides and eccentricity vectors of
both planets to vary, as well as the mass-ratio of the perturbing
planet. We assumed a plane-parallel configuration, and specified the
initial orbital elements at BJD$_{\mathrm{TDB}}$-2454833 = 130
days. We optimized the model using a Levenberg-Marquardt solver
\citep{Press:1993} with numerically-computed, double-sided
derivatives. We found relatively good fits near all of these
mean-motion resonances, with $\chi^2$ values ranging from 243 to 290
for 160 degrees of freedom. These chi-square values are uncomfortably
large, which may indicate that the timing uncertainties of the planets
are underestimated by 20\% (in the scenario in which the treatment of
correlated noise is not sufficient to account for this noise), or
created by an astrophysical source, both equally speculative with the
data we currently have. The mass ratios of the perturbing planets for
these solutions varied from $10^{-6}$ to $3\times 10^{-5}$
(\ref{tab:TTVFast}), while the eccentricities of both planets are
modest, $<0.15$. We leave a more detailed description of these results
to future work.

\begin{table}[ht!]
  \centering
  \caption{\label{tab:TTVFast} Numerical outcomes for the mass
    (perturber-to-star) and period (perturber-to-transiting planet)
    ratios for each one of the exterior perturber TTV solutions.}
  \begin{tabular}{c c c}
    \hline\hline
    Ratio & Mass-ratio               &  Period ratio \\
    \hline
    $<2:1$ & $1.87 \times 10^{-5}$ &  1.993\\
    $>3:1$ & $2.17 \times 10^{-5}$ & 3.009\\
    $<4:1$ & $1.18\times 10^{-5}$ & 3.991\\
    $<5:3$ & $2.06\times 10^{-6}$ & 1.666\\
    $>2:1$ & $2.63\times 10^{-5}$ & 2.007\\
    $<3:2$ & $1.23\times 10^{-6}$ & 1.498\\
    $>4:1$ & $1.23\times 10^{-5}$ & 4.006\\
    $>5:3$ & $2.96\times 10^{-6}$ & 1.667\\
    $<3:1$ & $2.31\times 10^{-5}$ & 2.994\\
    \hline
  \end{tabular}
\end{table}
  
\begin{figure}[ht!]
  \centering
  \includegraphics[width=.5\textwidth]{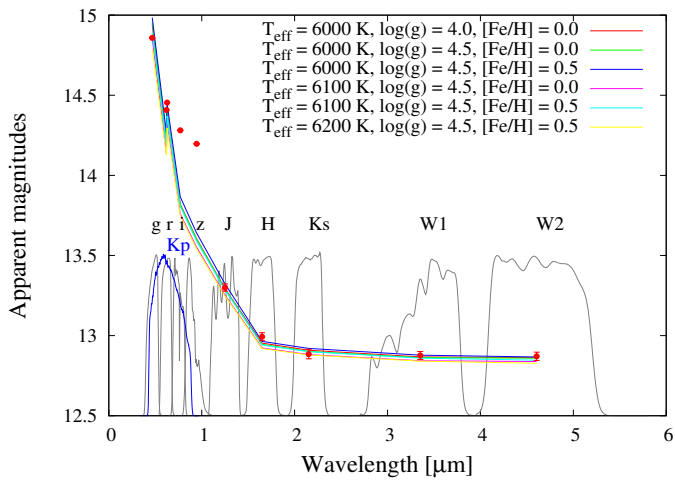}
  \caption{\label{fig:SED} Spectral energy distribution for
    KOI-0410. The figure shows apparent magnitudes as a function of
    wavelength, in $\mu$m. Shifted and scaled to match the figure, we
    have added in gray and blue continuous lines the filter
    transmission functions. On top of them, their respective names:
    Sloan g, r, i and z, Kepler band (K$_p$), 2MASS J, H and Ks, and
    WISE W1, W2. The first five quantities do not have uncertainties.}
\end{figure}

To investigate the gravitationally-bound hierarchical triple star
system scenario, we computed the Spectral Energy Distribution (SED) of
\mbox{KOI-0410}. For this end, we used all its available colors, taken
from the NASA Exoplanet Archive. These were compared to ``synthetic
colors'' which, in turn, were produced from PHOENIX stellar models. In
particular, when producing the synthetic colors we investigated a wide
range of stellar parameters, namely T$_{eff}$ = 6000, 6100, 6200
Kelvin, log(g) = 4.0, 4.5, and [Fe/H] = -0.5, 0.0. If more than one
star of similar spectral type would conform the \mbox{KOI-0410}
system, they should be revealed as an excess in the SED, when compared
to PHOENIX colors produced from a single star. We find a perfect match
between the observed and modeled colors for a star of \mbox{T$_{eff}$
  = 6100} Kelvin, \mbox{log(g) = 4.5}, and \mbox{[Fe/H] = -0.5}, close
to the values of \mbox{KOI-0410} reported in the bibliography, with
the exception of two photometric bands, namely i and z (see
Figure~\ref{fig:SED}), which we believe is caused by an incorrect
treatment of atmospheric extinction rather than astrophysics). Since
all the Sloan measurements do not have uncertainties, during the
upcoming 2018 observing season we will re-observe \mbox{KOI-0410} in
these photometric bands. At this point we would like to stress that,
for the \mbox{KOI-0410} system, the data available can not really
disentangle between the hierarchical triple star system and the
planet-star system, not even considering the TTVs. For example, if the
binary system was formed by two M-dwarf stars diluted by a G-type
star, then the stellar parameters and the radial velocity measurements
would be dominated by the latter, while the $\sim$14.4-day period
M-dwarf binary (diluted by the G-dwarf) might explain the Kepler
transits. A G-dwarf star orbiting an M-dwarf binary in a $\sim$3000
days, eccentric orbit might be sufficient to produce the observed TTVs
\citep{Borkovits2003,Agol2005}, causing an ambiguity with the
planet-star scenario. We leave a more detailed examination of this
scenario to future work.

\section{Conclusions}
\label{sec:CONCLUSIONS}

KOINet is a large collaboration spanning multiple telescopes across
the world aimed at achieving a follow-up coverage of KOIs exhibiting
TTVs. We have been focusing our instrumental capabilities initially on
60 KOIs that require additional data to complete a proper
characterization or confirmation by means of the TTV technique. A
complete list of these KOIs with all the relevant parameters has been
provided here.

There are several main challenges associated with the KOIs included in
this study: the faintness of their host stars, their shallowness and
their long duration. KOINet presents two fundamental advantages: the
access to {\it large telescopes} has allowed us to follow-up KOIs that
are faint ($\sim$13-16 K$_P$) and present shallow primary transit
events (1-10 ppt), minimizing observational biases. The advantage of
ground-based observations is the possibility to acquire short-cadence
data, of fundamental relevance for the determination of the
mid-transit times. Since the TTVs have already been detected by Kepler
and most of the systems show a TTV amplitude of several minutes,
detecting such offsets has been a straightforward task. The access to
{\it large longitudinal coverage} allowed us to have access to several
transit occurrences. In addition, since the transit duration grows
with the orbital period, for some of the KOIs the transit duration is
longer than the astronomical night at a given site. Therefore, more
than one site is required to fully observe the transit events. With
the observations collected during the 2014 and 2015 observing seasons
we have succeed with our timing precision requirements, we have added
new data improving the coverage of the TTV curves of systems where
Kepler did not register the interaction time fully, and we have built
a platform that can observe almost anywhere from the Northern
hemisphere. Although deriving planetary masses from transit timing
observations for more planets to populate the mass-radius-diagram is
an ambitious milestone, the work presented here shows we are on the
right track.

\begin{acknowledgements}

Funding for the Stellar Astrophysics Centre is provided by The Danish
National Research Foundation (Grant DNRF106). This research has made
use of the NASA Exoplanet Archive, which is operated by the California
Institute of Technology, under contract with the National Aeronautics
and Space Administration under the Exoplanet Exploration Program. E.A.
acknowledges support from NASA Grants NNX13A124G, NNX13AF62G, from
National Science Foundation (NSF) grant AST-1615315, and from NASA
Astrobiology Institutes Virtual Planetary Laboratory, supported by
NASA under cooperative agreement NNH05ZDA001C. S.W. acknowledges
support from the Research Council of Norway's grant 188910 to finance
service observing at the NOT, and support for International Team 265
(Magnetic Activity of M-type Dwarf Stars and the Influence on
Habitable Extra-solar Planets) funded by the International Space
Science Institute (ISSI) in Bern, Switzerland. J.F. acknowledges
funding from the German Research Foundation (DFG) through grant DR
281/30-1. Based on observations made with the Nordic Optical
Telescope, operated by the Nordic Optical Telescope Scientific
Association at the Observatorio del Roque de los Muchachos, La Palma,
Spain, of the Instituto de Astrof\'isica de Canarias. The data
presented here were obtained in part with ALFOSC, which is provided by
the Instituto de Astrof\'isica de Andaluc\'ia (IAA) under a joint
agreement with the University of Copenhagen and NOTSA. Based on
observations obtained with the Apache Point Observatory 3.5-meter
telescope, which is owned and operated by the Astrophysical Research
Consortium. Based on observations collected at the German-Spanish
Astronomical Center, Calar Alto, jointly operated by the
Max-Planck-Institut f\"ur Astronomie Heidelberg and the Instituto de
Astrof\'isica de Andaluc\'ia (CSIC). This work was supported in part
by the Ministry of Education and Science (the basic part of the State
assignment, RK no. AAAA-A17- 117030310283-7) and by the Act no. 211 of
the Government of the Russian Federation, agreement
no. 02.A03.21.0006. E.H. and I.R. acknowledge support by the Spanish
Ministry of Economy and Competitiveness (MINECO) and the Fondo Europeo
de Desarrollo Regional (FEDER) through grant ESP2016-80435-C2-1-R, as
well as the support of the Generalitat de Catalunya/CERCA
programme. E.P., G.T. and \v{S}.M.  acknowledges support from the
Research Council of Lithuania (LMT) through grant
LAT-08/2016. C.D.C. is supported by the Erasmus Mundus Joint Doctorate
Program by the grant number 2014-0707 from the EACEA of the European
Commission. S.E. acknowledges support by the Russian Science
Foundation grant No. 14-50-00043 for conducting photometric
observations of exoplanets of Kepler's mission. S.I. acknowledges
Russian Foundation for Basic Research (project No. 17-02-00542) for
support in the processing of the observations. K.P. acknowledges
support from the UK Science and Technology Facilities Council through
STFC grant ST/P000312/1. H.J.D. acknowledges support by grant
ESP2015-65712-C5-4-R of the Spanish Secretary of State for R\&D\&i
(MINECO). KOINet thanks the telescope operators for their valuable
help during some of the observing campaigns at the IAC80
telescope. The IAC80 telescope is operated on the island of Tenerife
by the Instituto de Astrof\'isica de Canarias in the Spanish
Observatorio del Teide. C.v.E. thanks the invaluable help and
contribution of all the telescope operators involved in this work.

\end{acknowledgements}

\bibliographystyle{aa} \bibliography{KOINet}

\end{document}